\documentclass[12pt,a4paper,final]{iopart}

\usepackage{iopams}  
\usepackage{graphicx}
\usepackage{subfigure}
\usepackage[breaklinks=true,colorlinks=true,linkcolor=blue,urlcolor=blue,citecolor=blue]{hyperref}

\expandafter\let\csname equation*\endcsname\relax
\expandafter\let\csname endequation*\endcsname\relax
\usepackage{amssymb}
\usepackage{amsmath}
\usepackage{amsfonts}
\usepackage{color}
\usepackage{mathrsfs}
\usepackage{caption}
\usepackage{float}

\begin{document}
\newtheorem{theorem}{Theorem}

\title[]{Shape, orientation and magnitude of the curl quantum flux, the
coherence and the statistical correlations in energy transport at nonequilibrium steady state}

\author{Zhedong Zhang$^{1}$ and Jin Wang$^{1,2,3}$}
\address{$^1$Department of Physics and Astronomy, SUNY Stony Brook, NY 11794, USA}
\address{$^2$Department of Chemistry, SUNY Stony Brook, NY 11794, USA}
\address{$^3$State Key Laboratory of Electroanalytical Chemistry, Changchun Institute of Applied Chemistry, Chinese Academy of Sciences, Changchun, Jilin, 130022, P. R. China}
\ead{jin.wang.1@stonybrook.edu}

\date{\today}

\begin{abstract}
We provide a quantitative description of the nonequilibriumness based on the model of coupled oscillators interacting with multiple energy sources. This can be applied to the study of vibrational energy transport in molecules. The curl quantum flux quantifying the nonequilibriumness and time-irreversibility is quantified in the coherent representation and we find the geometric description of the shape and polarization of the flux which provides the connection between the microscopic description of quantum nonequilibriumness and the macroscopic observables, i.e., correlation function. We use the Wilson loop integral to quantify the magnitude of curl flux, which is shown to be correlated to the correlation function as well. Coherence contribution is explicitly demonstrated to be {\it non-trivial} and to considerably promote the heat transport quantified by heat current and efficiency. 
 This comes from the fact that coherence effect is microscopically reflected by the geometric description of the flux. To uncover the effect of vibron-phonon coupling between the vibrational modes of molecular stretching (vibron) and the molecular chain (phonon), we further explore the influences of localization, which leads to the coherent and incoherent regimes that are characterised by the current-current correlations. 

\end{abstract}

\pacs{82.20.Rp, 88.20.jr}
\vspace{2pc}
\submitto{\NJP}

\section{Introduction}
Recently the nonequilibrium process exposes its significance in the study of microscopic properties and functions in molecules, i.e., the vibrational energy transport in protein and $\alpha$-helix \cite{Magana14,Leitner10,Lervik09,Lervik10,Leitner13,Pieniazek09,Brizhik06}. In the protein molecules surrounded by water, the vibrational energy transport is often mediated by the water molecules confined insides the protein, far less mobile than hydration water \cite{Hua06,Gnanasekaran09}. The vibrational energy transport in $\alpha$-helix is realized by the normal vibrational modes of the peptide group which consists essentially of the stretching of C=O bond \cite{Cruzeiro09}. Both of the circumstances, commonly, are affected by the strength of vibron-phonon interaction (i.e., hydrogen bond to system) between molecule stretching and the vibrations of molecular chain in a {\it non-trivial} manner \cite{Gnanasekaran09,Gnanasekaran11,Shenogina09}.

Various approaches have been developed to describe the vibrational energy transport in proteins \cite{Piazza09,Nguyen09,Yamato09,Backus08,Leitner08,Stock08}. The Davydov-Scott model \cite{Davydov73,Davydov91} assumed the systems are isolated from the external environments and worked under the mixed quantum-classical approximation \cite{Takeno97}, where the the quantum-mechanical wave functions of the excitations in amide I have to be determined and the motion of the peptide groups as a whole is treated classically. An alternative approach is the so-called frequency-resolved communication maps \cite{Gnanasekaran11,Leitner09,Xu14} developed in recent years, which locates the vibrational energy transfer channels in biomolecules. It is a coarse graining procedure to determine the thermal transport coefficients \cite{Leitner03} from the constraint on the size of the biomolecules. These approaches, however, are all based on the assumption of the equilibrium of the whole systems \cite{Hynes83,uzer91}. In fact the stationary energy transport in biomolecules shows an unidirectional feature, which in other words, indicates the time-irreversibility, since the energy is released from the chemical reactions and then transfered to the proteins. In this study we propose a theoretical framework based on nonequilibrium quantum process to describe the vibrational energy transport in biomolecules with the explicit detailed-balance-breaking, in the view of the quantum heat engine (QHE). Our main concern is on the coherence effect on the microscopic quantification of nonequilibriumness and also the macroscopic heat transfer properties.

The coherence mathematically dictated by the off-diagonal elements of density matrix has been explored in both theory and experiments at the nanoscale \cite{Engel07,Panitchayangkoon10,Engel12,Collini10,Plenio08,Nalbach15}, and found to facilitate the energy and charge transports in long distance. Recent investigations illustrate that the interaction between the excitations and discrete vibrational modes has the crucial contribution to the enhancement of coherence \cite{Renger01,Womick11,Zhang15}. Here in biomolecules, the vibrational modes refer to the motion of the protein peptides (i.e., hydrogen bond). To address such effect, we will adopt the conclusion as indicated by the previous studies \cite{Zhang15} that the excitation-vibration coupling leads to a renormalization to the detuning between the excitation frequencies and the coupling strengths between different excitations. 

In this article, we propose a theoretical framework for the description of the intrinsic nonequilibrium behaviors in quantum systems at steady state, based on the investigation of coupled harmonic oscillators connecting to multiple energy sources. This model can be used to study the vibrational energy transport in molecules surrounded by solvent. In analogous with the biochemical systems at classical level \cite{Wang08}, i.e., adenosine triphosphate hydrolysis (ATP) \cite{Xu12}, we analytically develope in coherent representation the curl quantum flux which quantifies the microscopic nonequilibriumness and the time-irreversibility \cite{Zhang14,Zhangnjp15}. The geometric description of the curl flux is found to uncover the connection between the microscopic description of quantum nonequilibriumness and the macroscopic correlation function. We further demonstrate that the coherence has {\it non-trivial} and considerable contribution to heat transfer, in the view of quantum heat engine (QHE). This is demonstrated by the comparison with the senario under secular approximation where the site coherence is decoupled from the population dynamics. This microscopically originates from the coherence effect on vibrational correlations reflected by the geometry of curl flux. It is much in slender-cigar shape orientated in the vicinity of anti-diagonal in phase space, as the coherence-population coupling increases. We use the Wilson loop integral to quantify the magnitude of curl flux, which is shown to be correlated to the correlation function as well. Finally we turn our attention to the influences of the localization, which aims at simulating the effect of vibron-phonon coupling. We particularly divide it into coherent and incoherent regimes, which are characterised by current-current correlation function.

\begin{figure}
\centering
 $\begin{array}{cc}
   \includegraphics[scale=0.55]{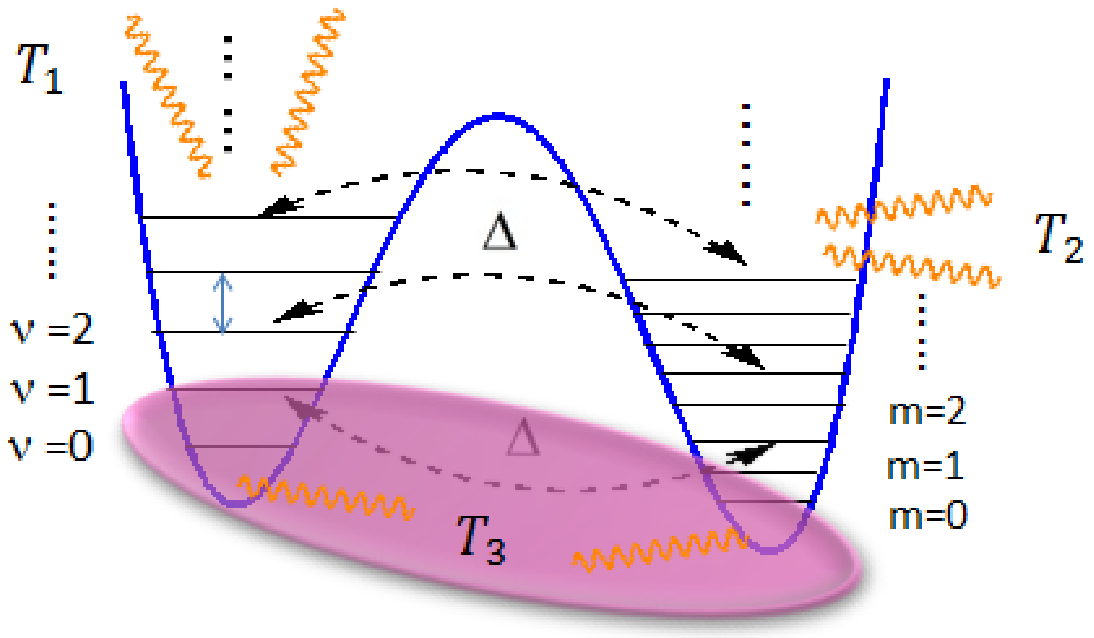}
   \includegraphics[scale=0.52]{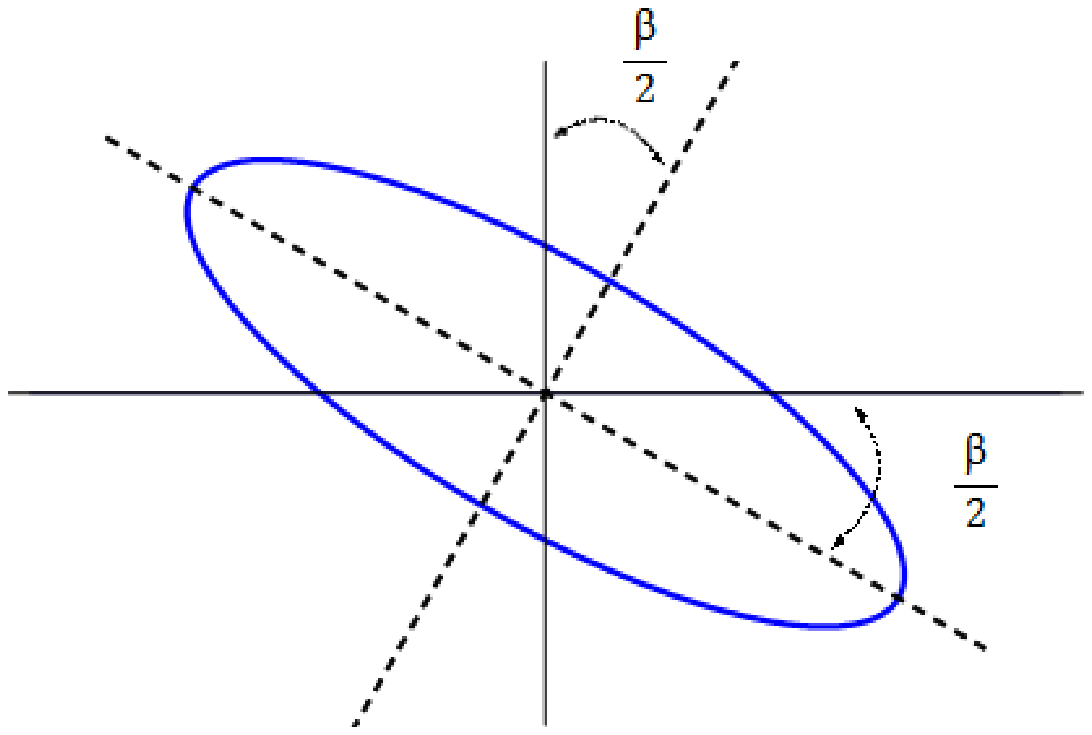}
  \end{array}$
\caption{(Color online) (Left) Schematic of vibrational energy transport in molecular system. Multi-level in each potential well represents the eigenstates of harmonic oscillator; (Right) Rotation of the ellipse to its principal axis.}
\label{schematic}
\end{figure}

\section{Hamiltonian and quantum master equation for nonequilibrium systems}
\subsection{Hamiltonian}
We consider the molecular vibrations (i.e., C=O stretching) described by two quantum-mechanically coupled oscillators with different frequency $\omega_1$ and $\omega_2$ which are immersed into the solvent environment at the interface. The free Hamiltonian for the system and solvent environment in terms of the displacements of oscillations reads
\begin{equation}
\begin{split}
H_{ssf} = \sum_{j=1}^2\left(\frac{p_j^2}{2m_j}+\frac{1}{2}m_j\omega_j^2 x_j^2\right) + \kappa & \left(\sqrt{m_1m_2}\omega_1\omega_2 x_1x_2+\frac{1}{\sqrt{m_1m_2}}p_1p_2\right)\\[0.2cm]
& + \sum_{\textbf{f},\sigma}\left(\frac{\tilde{p}_{\textbf{f}\sigma}^2}{2\tilde{m}_{\textbf{f}}}+\frac{1}{2}\tilde{m}_{\textbf{f}}\omega_{\textbf{f}\sigma}^2\tilde{x}_{\textbf{f}\sigma}^2\right)
\end{split}
\label{hs}
\end{equation}
where the low-energy fluctuation solvent environment is treated as a set of harmonic oscillators, neglecting the anharmonic effect hereafter. Actually this effect becomes important under some conditions which goes beyond the scope of this article and we would address this issue in the forthcoming studies. $m_j$ and $\tilde{m}_{\textbf{k}}$ are the effective masses of the molecular-vibrations and environmental vibration mode $\textbf{k}$, respectively; $p_j$ and $x_j$ are the canonical momentum and coordinate (describing the displacement of the oscillation), respectively. $\kappa$ is dimensionless that characterizes the coupling strength between two oscillators. In Eq.(\ref{hs}) we only pick up the minimal coupling between oscillation modes, in order to elucidate our model. It is in general replaced by the dipole-dipole interaction in many chemical systems. The interaction between the system and solvent environment is in similar form as that between the vibrations
\begin{equation}
\begin{split}
H_{int}^{ss} = \sum_{i=1}^2\sum_{\textbf{f},\sigma}\lambda_{\textbf{f}\sigma}^{ss}\left(\sqrt{m_i\tilde{m}_{\textbf{f}}}\omega_i\omega_{\textbf{f}\sigma}x_i \tilde{x}_{\textbf{f}\sigma}+\frac{1}{\sqrt{m_i\tilde{m}_{\textbf{f}}}}p_i \tilde{p}_{\textbf{f}\sigma}\right)
\end{split}
\label{ss}
\end{equation}

Additionally one oscillator interacts with a heat source characterizing the energy pump from the chemical reactions, and the other one interacts with a cooler environment which harvests the energy transfered by molecular vibrations, as shown in Fig.\ref{schematic}. The quantum-mechanically effective couplings of system to these reservoirs can be realized by exchanging the energy quanta. Generally the interaction between molecule stretchings is mediated via the dipole-dipole coupling, therefore in the formalism of {\it Quantum Field Theory}, the free Hamiltonian of system+reservoirs+environment is
\begin{equation}
\begin{split}
H_0 = \bar{\varepsilon}_1 a_1^{\dagger}a_1 + \bar{\varepsilon}_2 a_2^{\dagger}a_2 + \bar{\Delta}(a_1^{\dagger}a_2+a_2^{\dagger}a_1)+\sum_{\nu=1}^3\sum_{\textbf{k},\sigma}\hbar\omega_{\textbf{k}\sigma}b_{\textbf{k}\sigma}^{(\nu),\dagger}b_{\textbf{k}\sigma}^{(\nu)}
\end{split}
\label{1}
\end{equation}
and the interactions are
\begin{equation}
\begin{split}
H_{int} = \sum_{i=1}^2\sum_{\textbf{k},p}g_{\textbf{k}p}\left(a_i^{\dagger}b_{\textbf{k}p}^{(i)}+a_i b_{\textbf{k}p}^{(i),\dagger}\right) + \sum_{\textbf{q},\sigma}g_{\textbf{q}\sigma}\left(c^{\dagger}b_{\textbf{q}\sigma}^{(3)}+c\ b_{\textbf{q}\sigma}^{(3),\dagger}\right)
\end{split}
\label{2}
\end{equation}
where $c\equiv a_1+a_2$ and $b_{\textbf{k}\sigma}^{(1)}$, $b_{\textbf{k}\sigma}^{(2)}$ and $b_{\textbf{k}\sigma}^{(3)}$ are the bosonic annihilation operators for heat source, cool reservoir and solvent environment, respectively. $\sigma$ and $p$ denote the polarizations of bosons in the reservoirs. The rotating-wave approximation (RWA) \cite{Scully97,Breuer02} has been appiled to the vibration-bath interactions, owing to the dominant contribution by real absorption and emission of quanta in long time limit. The effect of vibron-phonon interaction (VP) can be approximately described by the renormalized frequency gap $\delta\bar{\varepsilon}=\bar{\varepsilon}_1-\bar{\varepsilon}_2$ and coupling strength $\bar{\Delta}$ of the molecular vibrations \cite{Zhang15}. This in particular, indicates that the tuning between frequencies $\delta\bar{\varepsilon}\ll \bar{\Delta}$ (strong VP) makes the wave packet of vibrations extended while the large detuning between frequencies $\delta\bar{\varepsilon}\gg \bar{\Delta}$ (weak VP) makes the wave packet localized, as elaborately illustrated in Anderson localization mechanism \cite{Anderson58,John91,Segev13}, in which the disorder (both diagonal and off-diagonal) is imaged to be connected with the presence of impurities, vacancies and dislocations in an ideal crystal lattice, or the random distributions of atoms and molecules \cite{Kramer93}.

\subsection{Quantum Master Equation in coherent space}
The dynamics of the system is governed by the reduced quantum master equation (RQME), which is obtained by tracing out the degree of freedoms (DOF) of the baths. As we pointed out above and also discussed in previous papers \cite{Zhang15,Fleming12,Scholes14}, the strong interactions between the system and some discrete vibrational modes (i.e., the hydrogen bond) owing to the quasi-resonance between frequencies, leads to the comparable time scales between the vibrational modes and system, which subsequently acquires us to include the dynamics of these discrete vibrational modes together with the system. In other words, these vibrational modes must be seperated from the reservoirs. Thereby the remaining modes consisting of the low-energy fluctuations can be reasonably treated as the baths, which are effectively in weak coupling to the systems due to the mismatch of frequencies between these continous modes and the system. On the basis of perturbation theory, the whole solution of the density operator can be written as $\rho_{SR}=\rho_s(t)\otimes\rho_R(0)+\rho_{\delta}(t)$ with the traceless term in higher orders of couplings between system and reservoirs. Because the time scale associated with the environmental correlations is much smaller than that of system over which the state varies appreciably, the RQME for the reduced density matrix of the systems in the interaction picture can be derived under the so-called Markoff approximation
\begin{equation}
\begin{split}
\frac{d\rho_s}{dt} = & \frac{1}{2\hbar^2}\bigg\{\sum_{\nu=1}^2\sum_{p=1}^2\left[\gamma_p^{T_{\nu},+}\left(a_p\rho_s a_{\nu}^{\dagger}-a_{\nu}^{\dagger}a_p\rho_s\right)+\gamma_p^{T_{\nu},-}\left(a_p^{\dagger}\rho_s a_{\nu}-a_{\nu}a_p^{\dagger}\rho_s\right)\right]\\
& \qquad + \sum_{j=1}^2\sum_{p=1}^2\left[\gamma_p^{T_3,+}\left(a_p\rho_s a_j^{\dagger}-a_j^{\dagger}a_p\rho_s\right) + \gamma_p^{T_3,-}\left(a_p^{\dagger}\rho_s a_j-a_ja_p^{\dagger}\rho_s\right)\right]\bigg\} + \textup{h.c.}
\end{split}
\label{3}
\end{equation}
where the reservoirs and solvent environment are in thermal equilibrium. The dissipation rates $\gamma_{...}$ are given in Appendix A. $n_{\omega}^T=\left[\textup{exp}(\hbar\omega/k_B T)-1\right]^{-1}$ is the Bose occupation on frequency $\omega$ at temperature $T$. The mixture angle reads
\begin{equation}
\begin{split}
\textup{cos}2\theta = \frac{\bar{\varepsilon}_2-\bar{\varepsilon}_1}{\sqrt{(\bar{\varepsilon}_1-\bar{\varepsilon}_2)^2+4\bar{\Delta}^2}};\quad \textup{sin}2\theta = -\frac{2\bar{\Delta}}{\sqrt{(\bar{\varepsilon}_1-\bar{\varepsilon}_2)^2+4\bar{\Delta}^2}}
\end{split}
\label{5}
\end{equation}

Conventionally the QME in Eq.(\ref{3}) was solved in Liouville space \cite{Esposito09,Esposito06}, by writing the density matrix as a supervector. This strategy however, seems to be unrealizable due to the infinite dimension of Fock space for bosons. Here we will solve the QME in the coherent representation, which was first developed by Glauber \cite{Glauber63}. It is alternatively named as {\it P}-representation, according to the terminology in quantum optics \cite{Scully97,Breuer02,Sudarshan63}. As is known, the eigenstate of the annihilation operators is $|\alpha_1,\alpha_2\rangle$ where the eigenequation $a_j|\alpha_1,\alpha_2\rangle=\alpha_j|\alpha_1,\alpha_2\rangle$ is satisfied. In terms of these components, the density matrix can be expanded into the following form \cite{Scully97}
\begin{equation}
\begin{split}
\rho_s(t) = \int P(\alpha_{\beta},\alpha_{\beta}^*,t)|\alpha_1,\alpha_2\rangle\langle\alpha_1,\alpha_2|d^2\alpha_1 d^2\alpha_2
\end{split}
\label{6}
\end{equation}
where $P(\alpha_{\beta},\alpha_{\beta}^*,t)$ is called the quasi-probability, due to the overcompleteness of the coherent basis and the non-positive definition of $P(\alpha_{\beta},\alpha_{\beta}^*,t)$. But this quasi-probability is always non-negative everywhere if the quantum system has classical analog \cite{Wolf95}. Using Eq.(\ref{6}) we can project the QME into coherent space and then obtain the following dynamical equation for $P(\alpha_{\beta},\alpha_{\beta}^*,t)$
\begin{equation}
\begin{split}
& \frac{\partial}{\partial t}P(\alpha_{\beta},\alpha_{\beta}^*) = \gamma \left[2\left(\frac{\partial}{\partial\alpha_1}\alpha_1+\frac{\partial}{\partial\alpha_2}\alpha_2\right)+\frac{\partial}{\partial\alpha_1}\alpha_2+\frac{\partial}{\partial\alpha_2}\alpha_1+\textup{c.c.}\right]P(\alpha_{\beta},\alpha_{\beta}^*)\\[0.2cm]
& \qquad + \gamma\left[2\textup{Y}_1^1\frac{\partial^2}{\partial\alpha_1^*\partial\alpha_1}+2\textup{Y}_2^2\frac{\partial^2}{\partial\alpha_2^*\partial\alpha_2}+\textup{Y}_{12}^{21}\left(\frac{\partial^2}{\partial\alpha_1^*\partial\alpha_2}+\frac{\partial^2}{\partial\alpha_1\partial\alpha_2^*}\right)\right]P(\alpha_{\beta},\alpha_{\beta}^*)
\end{split}
\label{7}
\end{equation}
where the coefficients $\textup{Y}_{...}$ are given in Appendix A. Eq.(\ref{7}) is of the same formalism as the classical Fokker-Planck equation \cite{Risken96}, apart from the complex variables. Moreover, this equation describes the Ornstein-Uhlenbeck process in the phase space, so that it is exactly solvable. Now we focus on the steady-state case with $t\rightarrow \infty$, and therefore the steady-state solution is of Guassian type
\begin{equation}
\begin{split}
P_{ss}(\alpha_{\beta},\alpha_{\beta}^*) = \frac{1}{Z}e^{-[a|\alpha_1|^2+b|\alpha_2|^2+2c\textup{Re}(\alpha_1^*\alpha_2)]}
\end{split}
\label{9}
\end{equation}
with the $a$, $b$ and $c$ being
\begin{equation}
\begin{split}
& a = \frac{4 \left(\textup{Y}_1^1+7\textup{Y}_2^2-2\textup{Y}_{12}^{21}\right)}{(\textup{Y}_1^1+\textup{Y}_2^2)^2+4\left[3\textup{Y}_1^1\textup{Y}_2^2-(\textup{Y}_{12}^{21})^2\right]},\ 
b = \frac{4 \left(7\textup{Y}_1^1+\textup{Y}_2^2-2\textup{Y}_{12}^{21}\right)}{(\textup{Y}_1^1+\textup{Y}_2^2)^2+4\left[3\textup{Y}_1^1\textup{Y}_2^2-(\textup{Y}_{12}^{21})^2\right]}\\[0.2cm]
& c = \frac{8 \left(\textup{Y}_1^1+\textup{Y}_2^2-2\textup{Y}_{12}^{21}\right)}{(\textup{Y}_1^1+\textup{Y}_2^2)^2+4\left[3\textup{Y}_1^1\textup{Y}_2^2-(\textup{Y}_{12}^{21})^2\right]},\ 
Z = \frac{\pi^2}{12}\Big\{(\textup{Y}_1^1+\textup{Y}_2^2)^2+4\left[3\textup{Y}_1^1\textup{Y}_2^2-(\textup{Y}_{12}^{21})^2\right]\Big\}
\end{split}
\label{10}
\end{equation}
In the forthcoming sections, we will discuss the nonequilibrium behaviors and heat transport based on this steady-state solution of quasi-probability in Eq.(\ref{9}) and (\ref{10}).

\section{Shape and orientation of the curl flux and vibration correlations in non-equilibrium quantum systems}
\subsection{Shape and orientation of curl flux}
The quantitative description of nonequilibriumness, especially the far-from-equilibrium case in quantum systems was devoid, although it is known that macroscopical current would be observed if the system deviates from equilibrium. Recently the curl flux in discrete space was developed to describe the nonequilibrium behaviors in quantum systems \cite{Zhang14,Zhangnjp15}. Here we will alternatively develope a curl quantum flux in the continous space, in analogous with the classical case \cite{Wang08}. We are able to write the probabilistic evolution of diffusion equation (\ref{7}) covering the whole coherent space into the form of $\partial_t P+\nabla_{\alpha}\cdot\textbf{J}=0$ which gives the steady-state case $\nabla_{\alpha}\cdot\textbf{J}=0$. For the nonequilibrium systems at steady state in general, this does not necessarily mean that the flux $\textbf{J}$ has to vanish, due to the detailed-balance-breaking. Instead, the divergence-free nature implies that the flux is a rotational curl field in coherent space. To further elucidate the nonequilibrium nature of the curl quantum flux, we will reduce our discussion into the space spanned by ($x_1,x_2$) where $x_j=\textup{Re}[\alpha_j]$. On the other hand, the issue of even and odd variables then does not arise, because of the time-reversibility of $x_j$. 

Assuming $\alpha_j=x_j+ip_j$ where $x_j=\sqrt{2m_j\omega_j/\hbar}\langle\alpha_1,\alpha_2|\hat{x}_j|\alpha_1,\alpha_2\rangle$ are the dimensionless means of the displacements of the oscillators in coherent space, the dynamical equation of probability in ($x_1,x_2$) space becomes
\begin{equation}
\begin{split}
\frac{\partial}{\partial t}P(x_1,x_2) = \gamma & \left(2\frac{\partial}{\partial x_1}x_1+2\frac{\partial}{\partial x_2}x_2+\frac{\partial}{\partial x_1}x_2+\frac{\partial}{\partial x_2}x_1\right)P(x_1,x_2)\\[0.2cm]
& \qquad\qquad\quad +\frac{\gamma}{2}\left(\textup{Y}_1^1\frac{\partial^2}{\partial x_1^2}+\textup{Y}_2^2\frac{\partial^2}{\partial x_2^2}+\textup{Y}_{12}^{21}\frac{\partial^2}{\partial x_1\partial x_2}\right)P(x_1,x_2)
\end{split}
\label{11}
\end{equation}
the steady-state solution to which reads
\begin{equation}
\begin{split}
P_{ss}(x_1,x_2) = \frac{\sqrt{ab-c^2}}{\pi}e^{-(a x_1^2+b x_2^2+2c x_1x_2)}
\end{split}
\label{12}
\end{equation}
Thus the curl quantum flux is of the form
\begin{equation}
\begin{split}
& \mathcal{J}_{x_1} = \frac{\gamma\left(\textup{Y}_1^1-\textup{Y}_2^2\right)P_{ss}}{(\textup{Y}_1^1+\textup{Y}_2^2)^2+4\left[3\textup{Y}_1^1\textup{Y}_2^2-(\textup{Y}_{12}^{21})^2\right]}\\[0.2cm]
& \qquad\qquad\qquad \times\left[2\left(\textup{Y}_1^1+\textup{Y}_2^2-2\textup{Y}_{12}^{21}\right)x_1 + \left(7\textup{Y}_1^1+\textup{Y}_2^2-2\textup{Y}_{12}^{21}\right)x_2\right]\\[0.3cm]
& \mathcal{J}_{x_2} = -\frac{\gamma\left(\textup{Y}_1^1-\textup{Y}_2^2\right)P_{ss}}{(\textup{Y}_1^1+\textup{Y}_2^2)^2+4\left[3\textup{Y}_1^1\textup{Y}_2^2-(\textup{Y}_{12}^{21})^2\right]}\\[0.2cm]
& \qquad\quad\qquad \times\left[\left(\textup{Y}_1^1+7\textup{Y}_2^2-2\textup{Y}_{12}^{21}\right)x_1 + 2\left(\textup{Y}_1^1+\textup{Y}_2^2-2\textup{Y}_{12}^{21}\right)x_2\right]
\end{split}
\label{13}
\end{equation}

The behaviors of curl quantum flux are illustrated in Fig.\ref{flux}, which provides a description of nonequilibriumness on microscopic level. It is straightforward to prove that the stream lines of the curl flux in 2D space in our case form a set of ellipses, with the major axis in the vicinity of the anti-diagonal line. Therefore the polarization of the curl flux can be quantified in terms of geometric language, namely the {\it eccentricity} $\bar{e}$ and {\it rotation angle} $\beta/2$
\begin{equation}
\begin{split}
\bar{e} = \sqrt{\frac{2\sqrt{(a-b)^2+4c^2}}{a+b+\sqrt{(a-b)^2+4c^2}}},\quad \textup{sin}\beta = -\frac{2c}{\sqrt{(a-b)^2+4c^2}}
\end{split}
\label{geo}
\end{equation}
In particular, based on Eq.(\ref{12}) and (\ref{13}) we know that the shape and orientation of curl flux is governed by the eccentricity $\bar{e}$ and rotation angle $\beta/2$ where the polarization is in slender-cigar shape along the vicinity of the line $|x_1|=|x_2|$ as $\bar{e}$ increases and $\beta/2$ aprroaches $\pi/4$.

Fig.\ref{flux}(a) and \ref{flux}(b) show the effect of vibron-phonon coupling (VP) on the curl flux with the eccentricities $\bar{e}=0.978$ and $\bar{e}=0.957$, respectively. As we can see, the strong-VP-bond-contributed delocalization of the excited vibrational modes causes the flux to be more polarized in the vicinity of anti-diagonal than the weak-VP-bond-contributed 

\begin{figure}[H]
\centering
 $\begin{array}{cc}
  \includegraphics[scale=0.5]{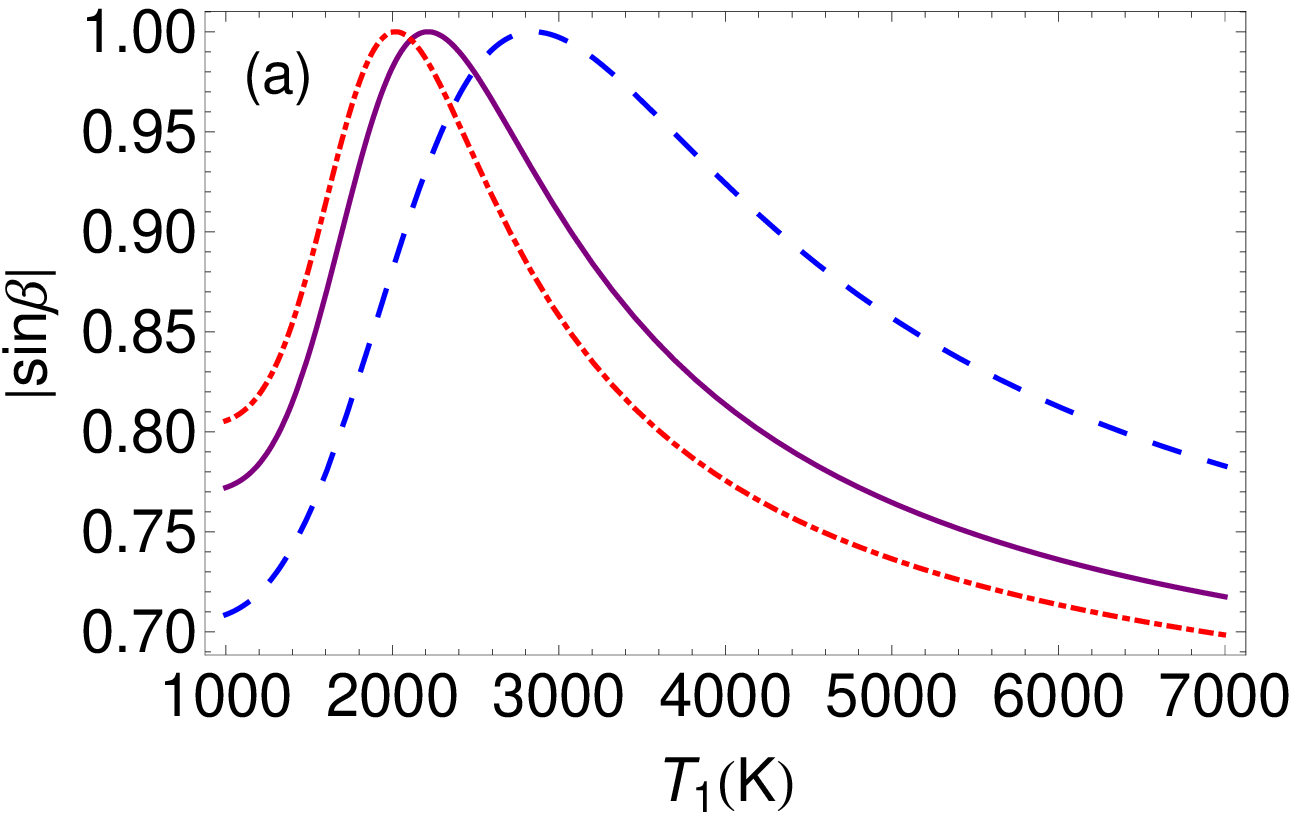}
 &\includegraphics[scale=0.52]{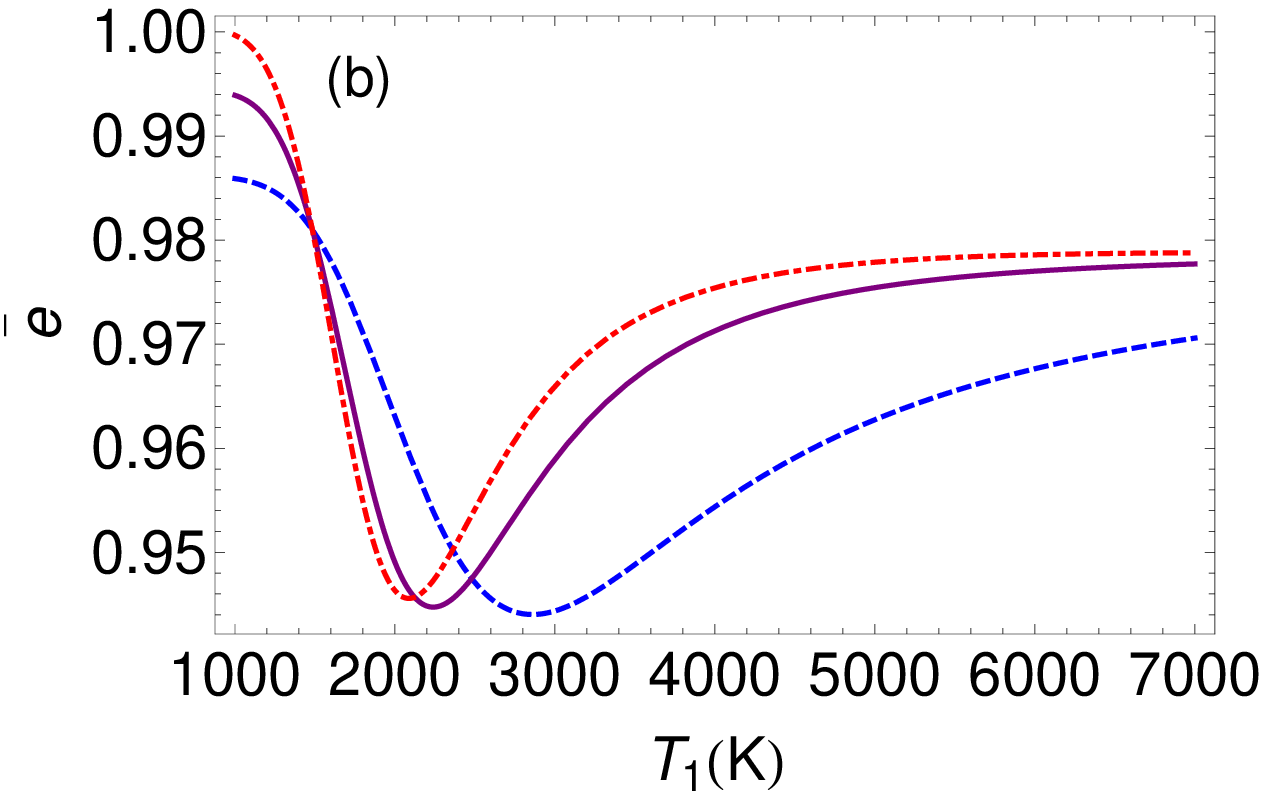}\\
  \includegraphics[scale=0.5]{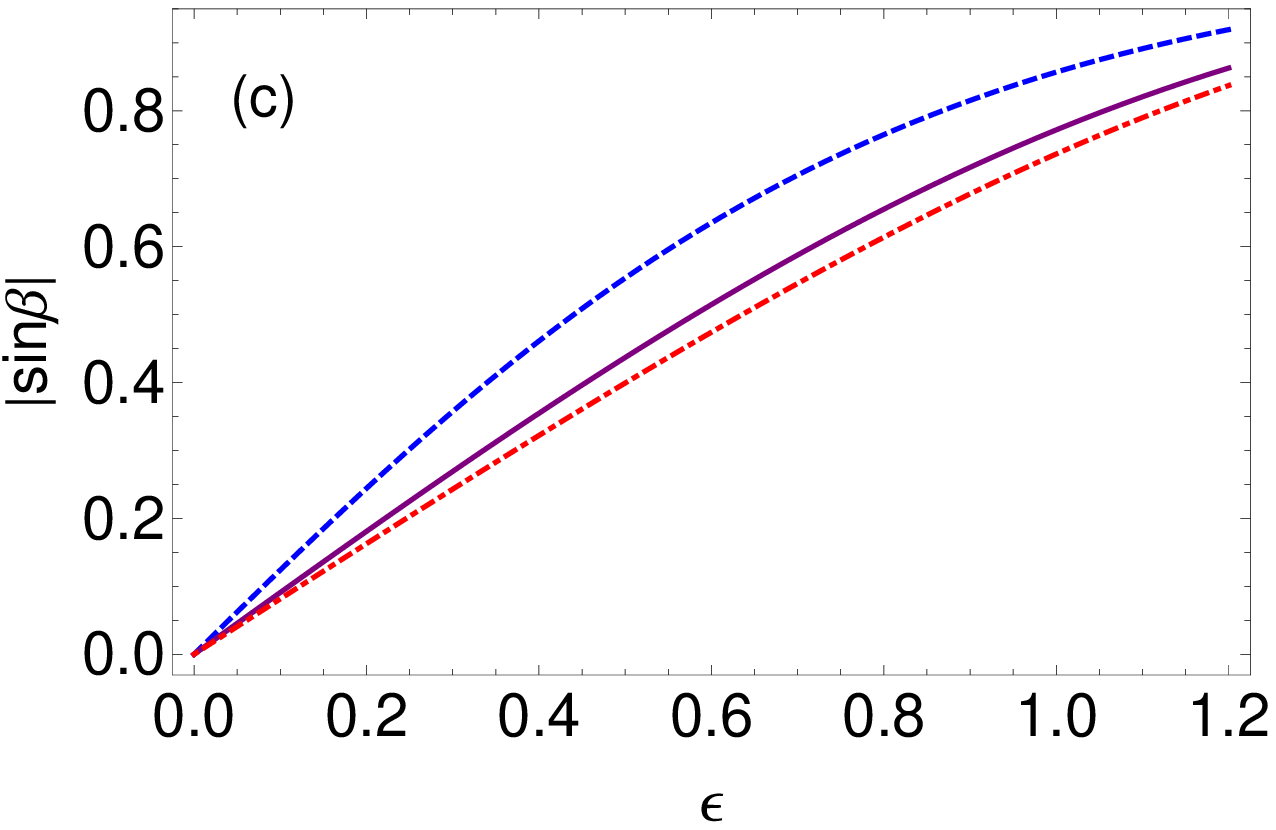}
 &\includegraphics[scale=0.5]{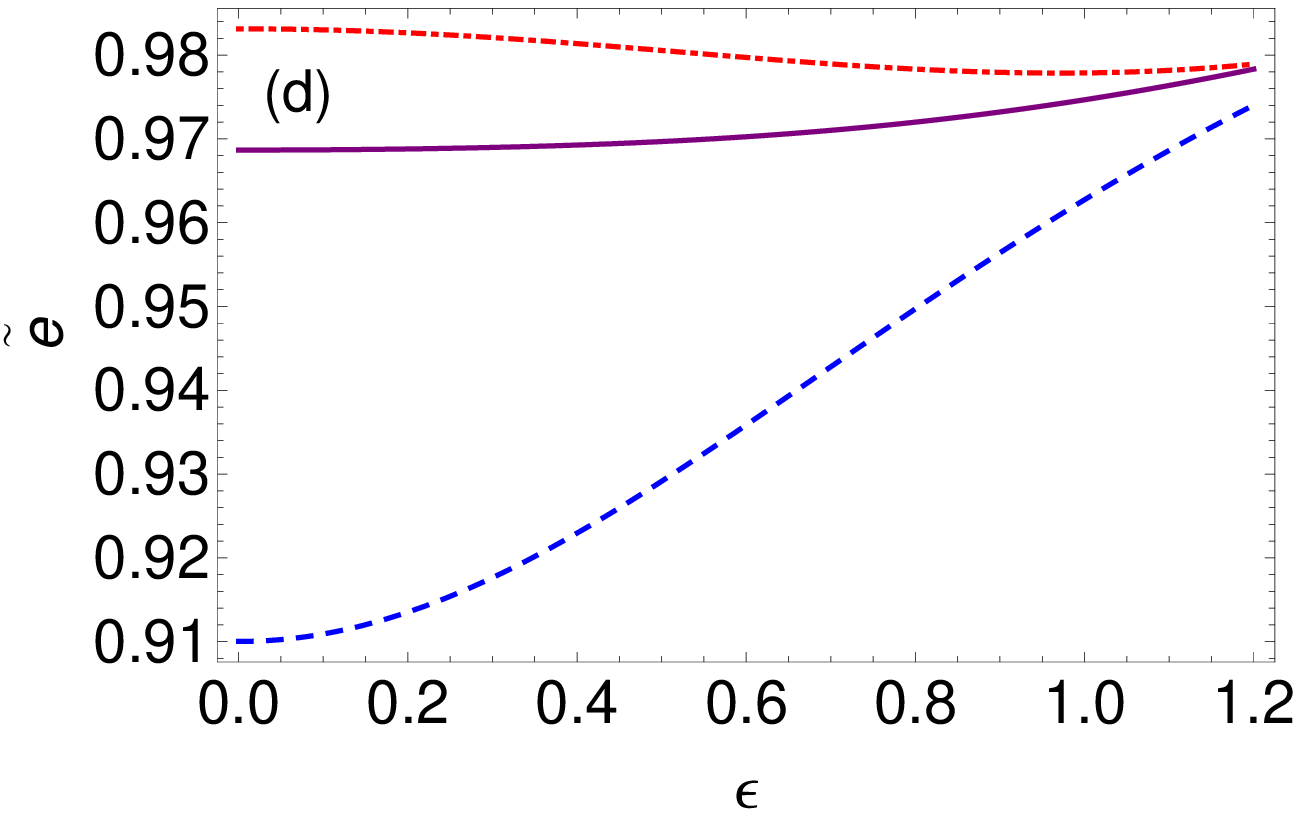}\\
  \includegraphics[scale=0.475]{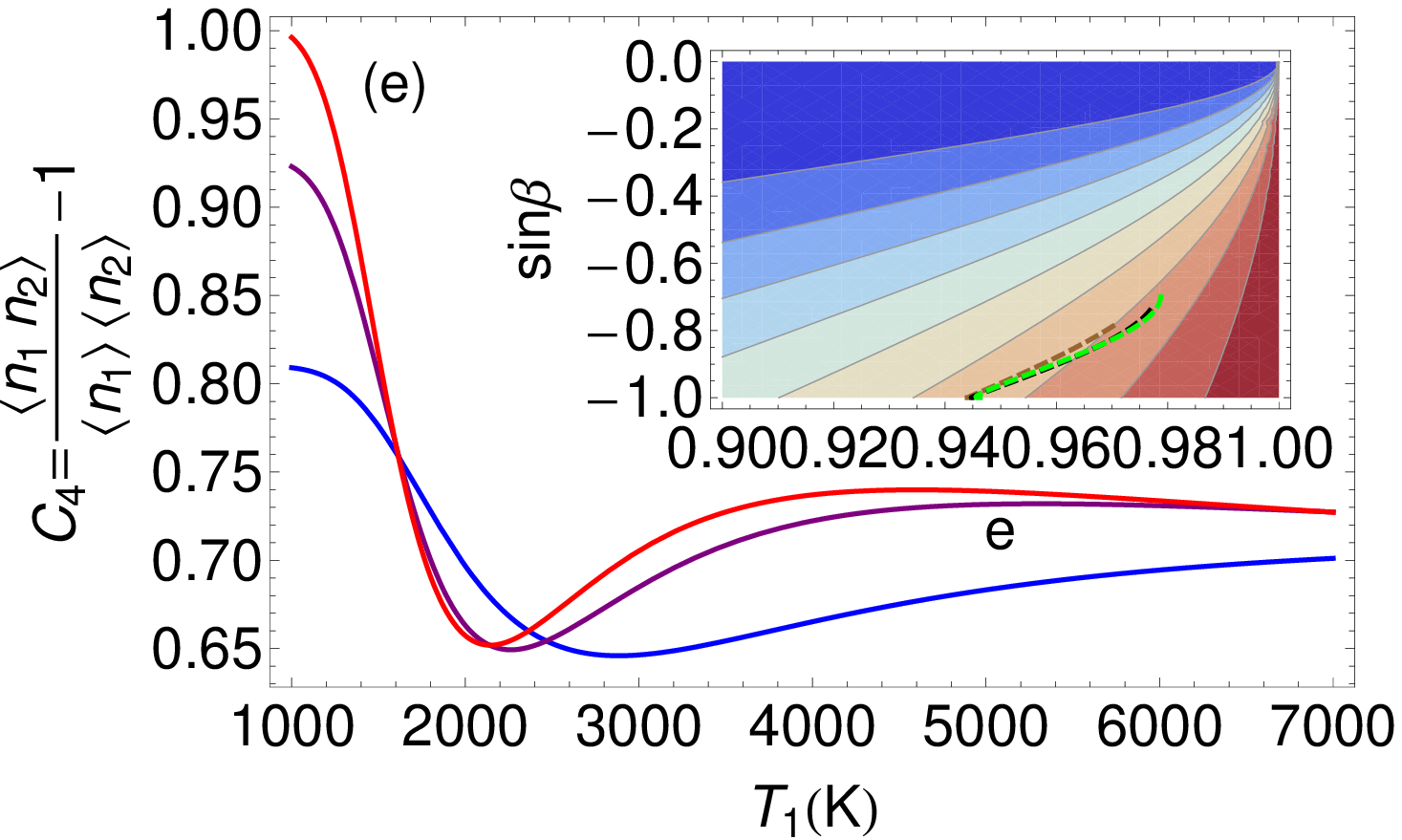}
 &\includegraphics[scale=0.51]{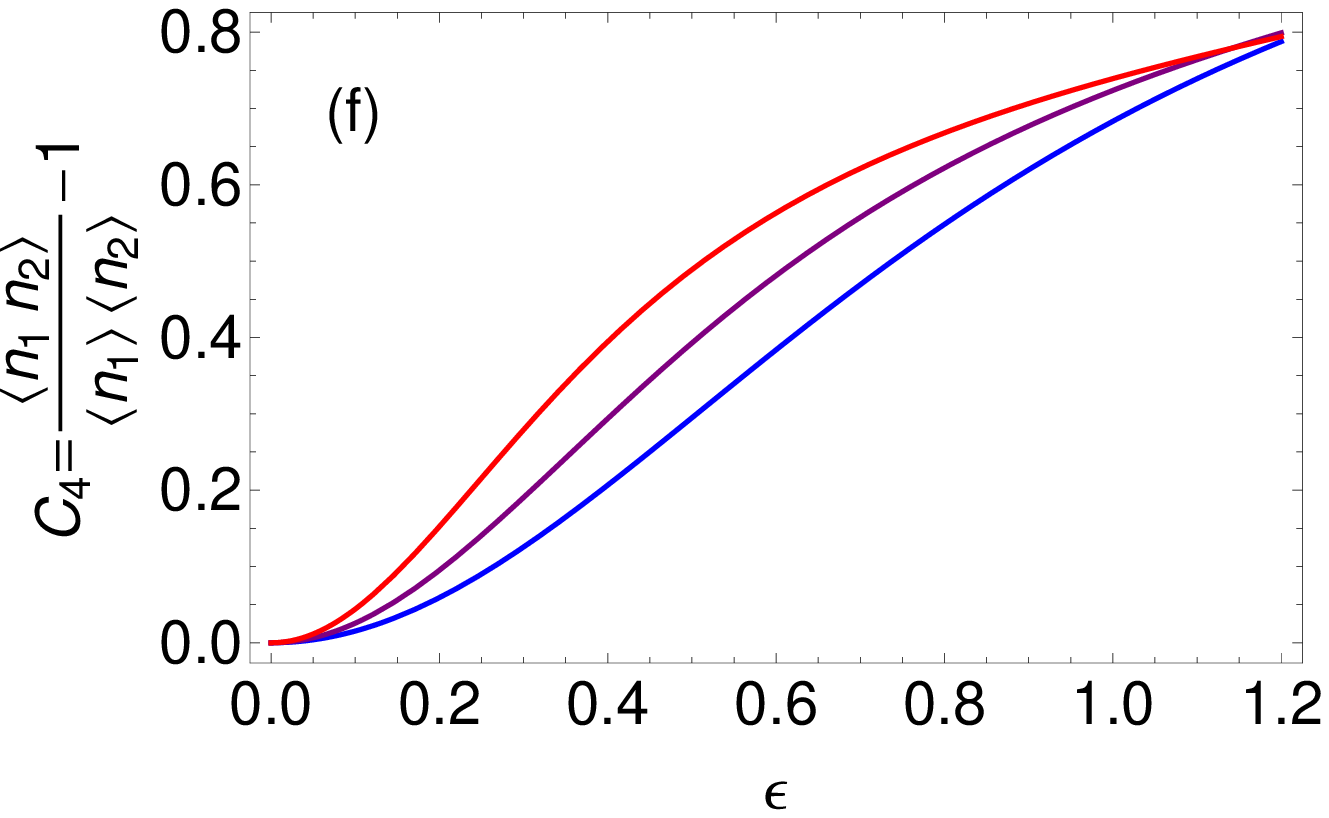}\\
  \includegraphics[scale=0.55]{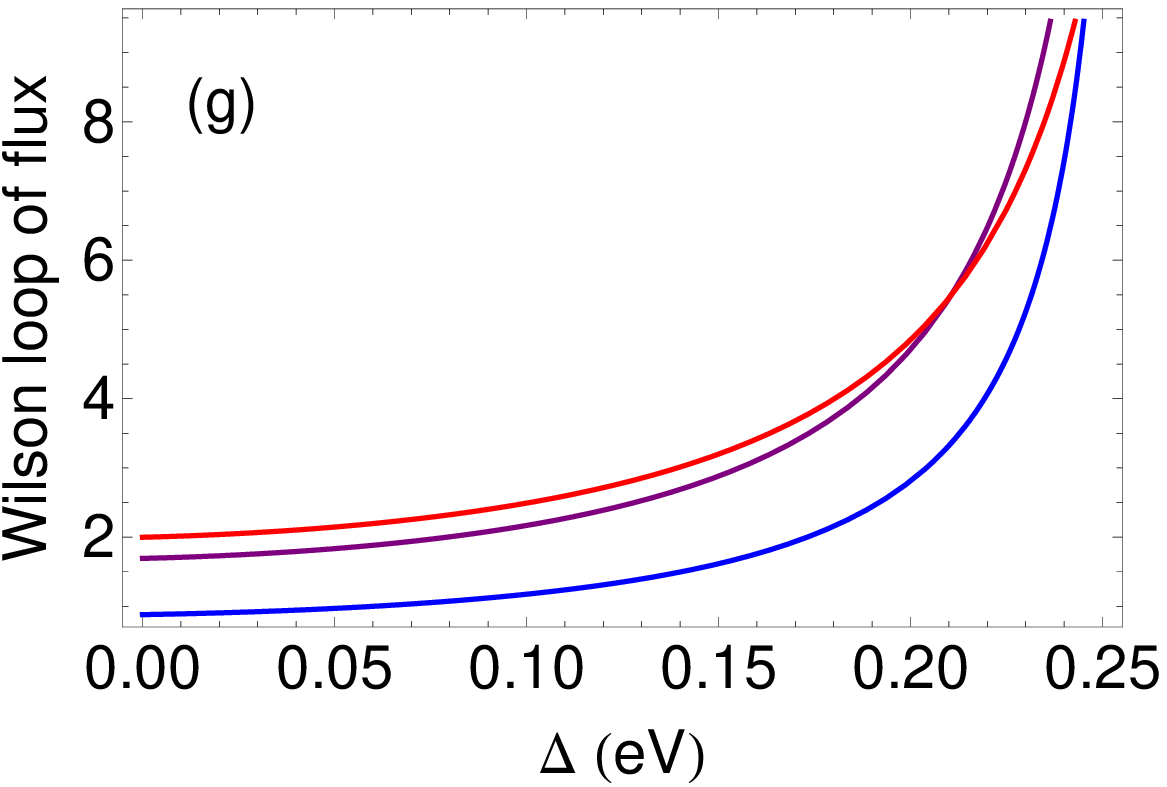}
 &\includegraphics[scale=0.54]{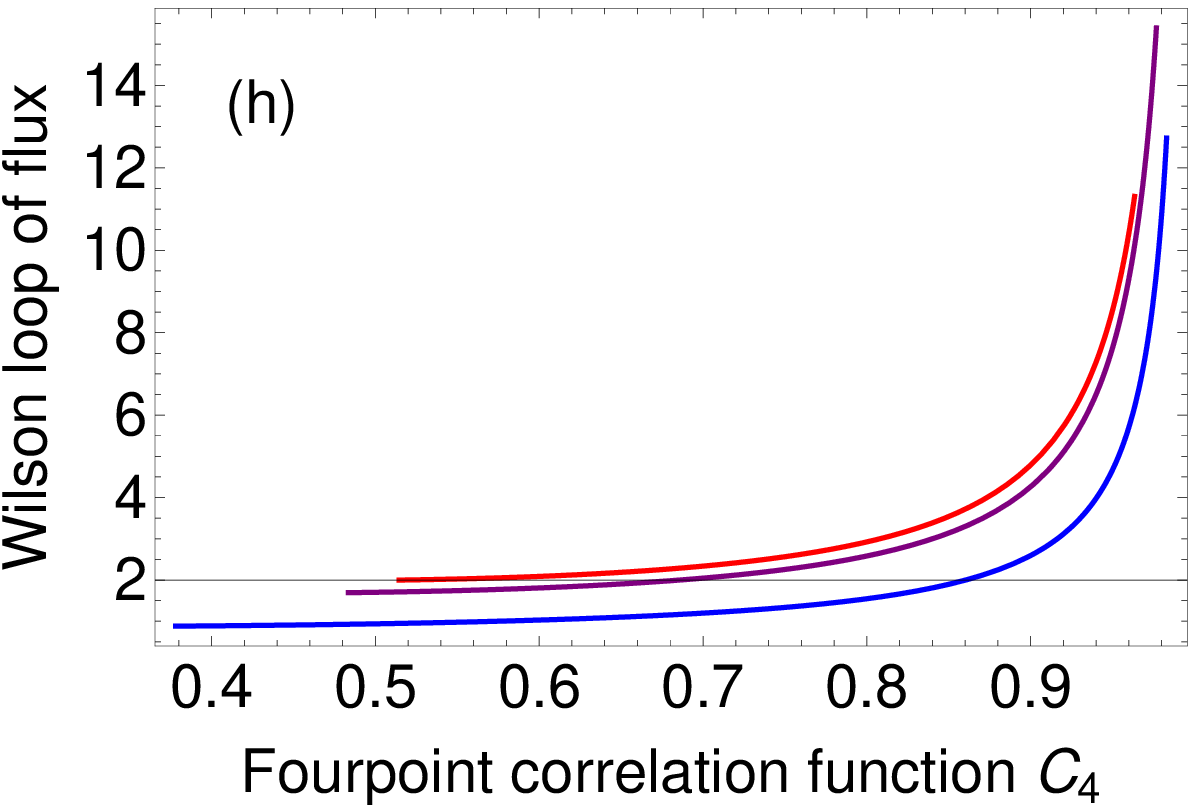}\\
  \end{array}$
\caption{(Color online) (a) Rotation angle and (b) eccentricity as a function of $T_1$; (c) Rotation angle and (d) eccentricity as a function of the coherence-population entanglement; (e) density-density correlation $C_4$ (large) via $T_1$ as well as 3D illustration of $C_4$ via eccentricity and rotation angle (small) according to Eq.(\ref{cor}), (f) density-density correlation vary as a function of the coupling strength between coherence and population dynamics; (g) Wilson loop of curl flux as a function of $\bar{\Delta}$; (h) Relationship between Wilson loop of curl flux and $C_4$ by controlling $\bar{\Delta}$. The blue (dashed), purple (solid) and red (dotdashed) lines in correspond to $\delta\bar{\varepsilon}=0.3,\ 0.1,\ 0.01$eV, respectively; In (c,d,g,h) $T_1=5000$K. Other parameters are $\bar{\Delta}=0.1$eV, $T_2=2000$K and $T_3=1000$K.}
\label{correlation}
\end{figure}
\noindent localization does, which as explored later, means that the correlation between the molecular vibrations is much stronger as quasi-particles become delocalized, rather than the localization of quasi-particles. The quality of heat transport will be promoted as the correlation between vibrations becomes strong, as shown in the following discussion. Fig.\ref{flux}(c) and \ref{flux}(d) show that the thermal fluctuations in heat source which in some sense dictates the effective thermal voltage, can definitely strengthen the molecular-vibration correlation, which will considerably raise the heat transport, as will be shown in Fig.\ref{heat}(c). Fig.\ref{flux}(e) and \ref{flux}(f) illustrate the effect of solvent environment on the curl flux, which shows that the correlation between molecular vibrations is unavoidably suppressed by the thermal fluctuations induced by solvent environment.

\subsection{Measure of magnitude of curl flux}
Due to the vector feature of the curl flux, here we will use the Wilson loop to quantify
the value of the curl flux, based on the integral of curl flux along a specific closed path
\begin{equation}
\begin{split}
W = \frac{1}{L}\int_{\Sigma}J_{x_1}\text{d}x_1+J_{x_2}\text{d}x_2
\end{split}
\label{loop}
\end{equation}
where $L$ represents the length of closed path $\Sigma$. Currently we choose the closed path as one of
the stream lines of curl flux in 2D space, to maximize the quantity $W$ in Eq.(\ref{loop}). After some manipulations the Wilson loop of curl flux reads
\begin{equation}
\begin{split}
W = \frac{\gamma|\text{Y}_1^1-\text{Y}_2^2|(a+b)}{16\text{E}(\frac{\pi}{2}|\bar{e}^2)\sqrt{e}}\sqrt{\frac{ab-c^2}{a+b+\sqrt{(a-b)^2+4c^2}}}
\end{split}
\label{W}
\end{equation}
Notice that $e=2.71828...$ is the Euler's number and $\text{E}(\phi|k^2)$ is the elliptic integral of the second kind. As is shown in Fig.\ref{correlation}(g), the detailed-balance is more broken as the two molecules approaches each other, governed by the increase of  $\frac{\bar{\Delta}}{\delta\bar{\varepsilon}}$. This can be understood by an extreme case that the two molecular vibrations will equilibrate individually with the environments as they becomes infinitely distanced with no correlation between each other ($\bar{\Delta}=0$). 

\begin{figure}[H]
\centering
 $\begin{array}{c}
  \includegraphics[scale=0.6]{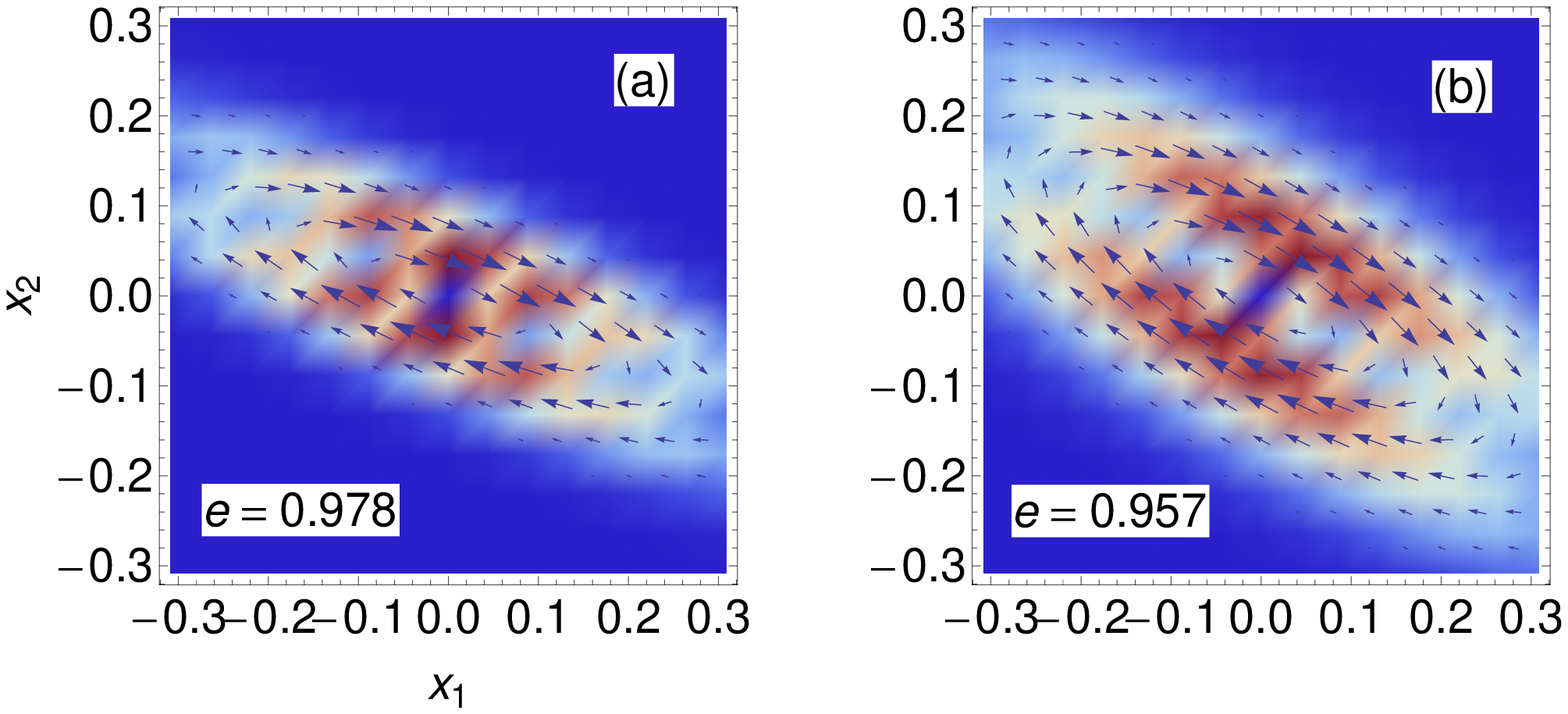}\\
  \includegraphics[scale=0.62]{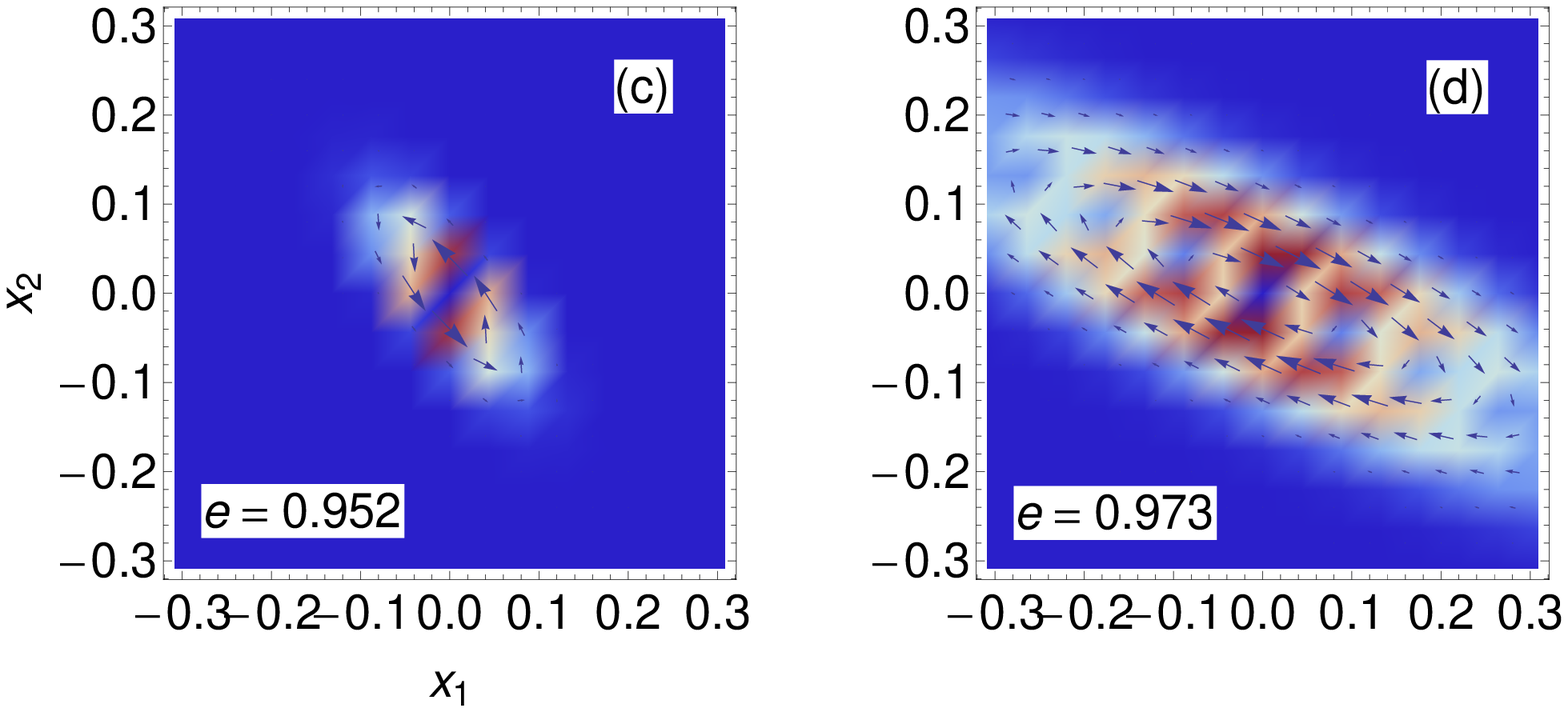}\\
  \includegraphics[scale=0.6]{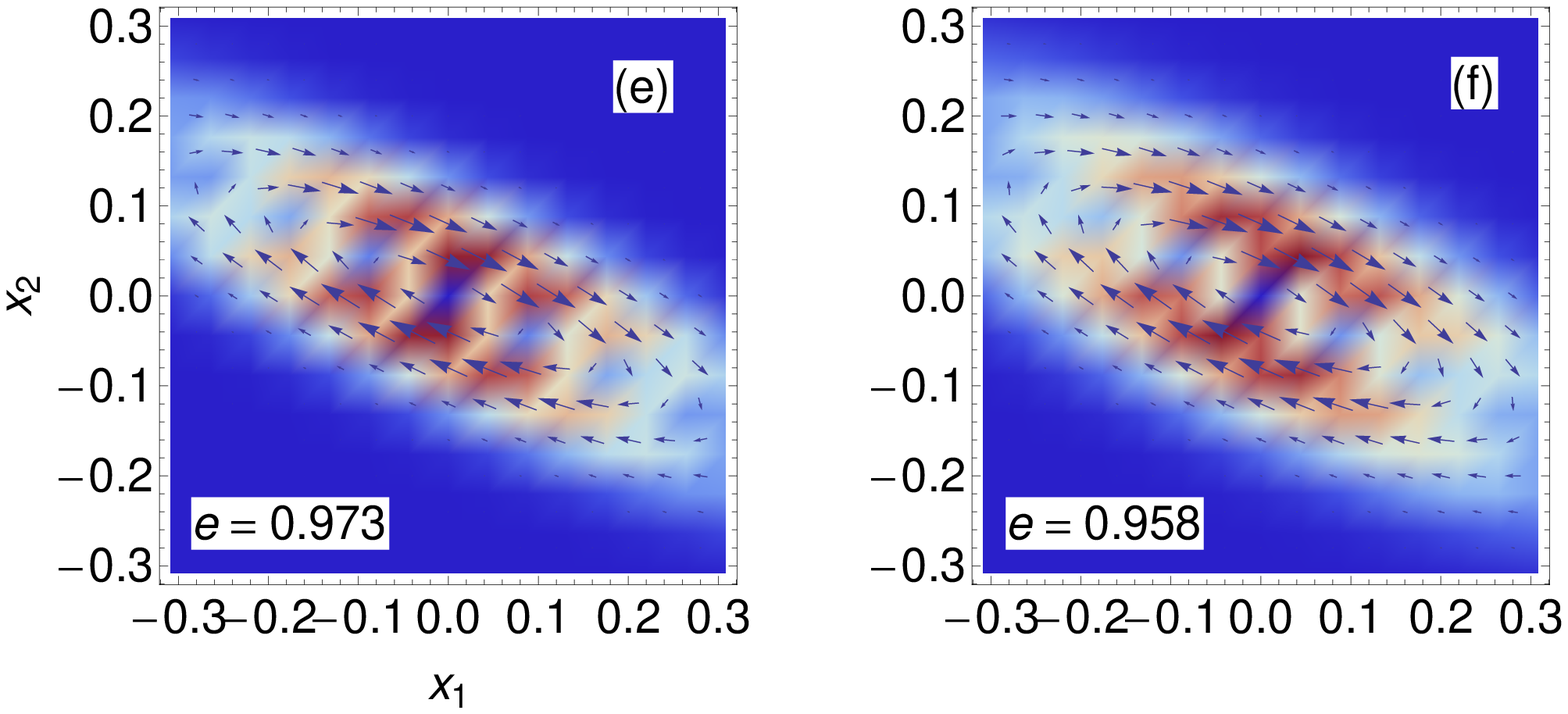}\\
  \end{array}$
\caption{(Color online) 2D Illustration of curl quantum flux. (a) $\bar{\Delta}=0.1\textup{eV},\ \delta\bar{\varepsilon}=0.01\textup{eV}$ and (b) $\bar{\Delta}=0.1\textup{eV},\ \delta\bar{\varepsilon}=0.35\textup{eV}$ where $T_1=5000$K and $T_3=1000$K; (c) $T_1=2000$K, $T_3=1000$K and (d) $T_1=5000$K, $T_3=1000$K where $\bar{\Delta}=0.1\textup{eV},\ \delta\bar{\varepsilon}=0.15\textup{eV}$; (e) $T_1=5000$K, $T_3=1000$K and (f) $T_1=5000$K, $T_3=2000$K where $\bar{\Delta}=0.1\textup{eV},\ \delta\bar{\varepsilon}=0.15\textup{eV}$. Other parameters are $\bar{\varepsilon}_1=1$eV and $T_2=2000$K. Notice that (d) and (e) are the same. However, the reason why we keep (e) here is to provide a control to (f), showing the effect of solvent. In these figures, the eccentricity is denoted by $e$.}
\label{flux}
\end{figure}

\subsection{Relationship between curl flux and vibration correlation}
The correlations between the molecular vibrations are measured by the correlation functions. In particular, we mainly focus on the four-point correlation functions here, which corresponds to the density-density correlations
\begin{equation}
\begin{split}
C_4 = \frac{\langle a_1^{\dagger}a_1a_2^{\dagger}a_2\rangle}{\langle a_1^{\dagger}a_1\rangle\langle a_2^{\dagger}a_2\rangle}-1=
\frac{\bar{e}^4\textup{sin}^2\beta}{4(1-\bar{e}^2)+\bar{e}^4\textup{sin}^2\beta}
\end{split}
\label{cor}
\end{equation}
where $\bar{e}$ shares the same definition of the eccentricity as before, and $\beta/2$ is the angle between the major axis of the ellipse and $x$-axis. 
Eq.(\ref{cor}) follows the definition of density-density correlation function in the site basis, as given in Ref.\cite{Scully97}. $C_4=0$ indicates that the occupations on the two vibrational modes are uncorrelated. Eq.(\ref{cor}) in fact uncovers the connection between the microscopic nonequilibriumness (flux) and the macroscopic observables in a geometric manner. Hence it is evident to say that {\it the correlations between molecular vibrations are charaterised by the polarization of the curl flux in coherent space, where the slender-cigar type of polarization of flux along the anti-diagonal means the strong vibrational correlations while the cake type of polarization with tiny inhomogeneity or polarization along the axis means the weak vibrational correlations.} Possibly the microscopic curl quantum flux can be explored in experiments, by measuring these geometric parameters through the measurement of density-density correlation function, which has been recently probed in degenerate quantum gas \cite{Jin05,Manz10}.

The large figure in Fig.\ref{correlation}(e) show that (i) the thermal fluctuations in the heat source do strengthen the vibration correlations and (ii) the detuning between the frequencies of molecular vibrations distroys the vibration correlations, as long as the heat transport is on track (this will be examined in detail in the section of heat transport). These properties are also microscopically reflected by curl flux, as shown in Fig.\ref{correlation}(a), Fig.\ref{correlation}(b) and the paths (dashed lines) on the landscape of density-density correlation $C_4$ in the small figure in Fig.\ref{correlation}(e), in spite of the reduction of the contribution by rotation angle. 
Besides the geometry of the curl flux, the promotion of the magnitude of curl flux (quantified by Wilson loop) is also strongly correlated to the increase of correlation function, by improving the coupling strength between vibrational modes, as shown in Fig.\ref{correlation}(h). 

As will show later in the section of coherence effect, the site-basis coherence has no contribution to the heat transport in the secular approximation since it is decoupled from the population dynamics. As approaching this regime, one can easily demonstrate that $\lim\limits_{\epsilon\rightarrow 0}\tilde{c} =0$ in Eq.(\ref{27}), which subsequently gives $\beta \rightarrow 0$. Hence $\lim\limits_{\epsilon\rightarrow 0}C_4 = 0$, which is physically reasonable owing to the polarization of flux at the moment orientatied in the vicinity of $x_1$-axis.
 As the entanglement between coherence and population dynamics adiabatically increases, the macriscopic correlation $C_4$ is considerably promoted by coherence from the microscopic curl flux as it becomes significantly polarized with the angle approaching $-\frac{\pi}{4}$, as illustrated in Fig.\ref{correlation}(c,d) and Fig.\ref{correlation}(f). Therefore we can conclude that the coherence generates and  considerably improves the correlation between the molecular vibrations.

\section{Heat transport intermediated by molecular vibrations}
To uncover the behaviors of vibrational energy transport in the molecules, one essentially needs to study the macroscopic heat current flowing through the molecular chain, output work in the view of the system as a quantum heat engine (QHE) and the efficiency quantifying the quality of this QHE. On the other hand, the correlation between the heat currents are also useful to measure to heat transport.
\subsection{Heat current and working efficiency}
We first introduce the heat-current operators $J_1$, $J_2$ for the heat currents pumping into and flowing out from the system, respectively. By ignoring the back influence of system to reservoirs, the current operators are defined as $J_{\nu}=\frac{1}{i\hbar}[H_s,H_{int}^{(\nu)}]$, and furthermore in our system
\begin{equation}
\begin{split}
J_1 = \frac{1}{i\hbar}\sum_{\textbf{k},\sigma}g_{\textbf{k}\sigma}\left[\left(\bar{\varepsilon}_1a_1^{\dagger}+\bar{\Delta}a_2^{\dagger}\right)b_{\textbf{k}\sigma}^{(1)}-\textup{h.c.}\right], J_2 = \frac{i}{\hbar}\sum_{\textbf{k},\sigma}g_{\textbf{k}\sigma}\left[\left(\bar{\varepsilon}_2a_2^{\dagger}+\bar{\Delta}a_1^{\dagger}\right)b_{\textbf{k}\sigma}^{(2)}-\textup{h.c.}\right]
\end{split}
\label{14}
\end{equation}
To calculate the heat current up to the 2nd order of coupling strength, we need to carry out the 1st order correction to the density matrix so that $\rho_s(t) = \rho_s(0)+\frac{i}{\hbar}\int_0^t[\rho_s(t),\tilde{H}_{int}(\tau)]d\tau$.
 Substituting this into the product $\rho_s(t)\tilde{J}_1(t)$ and after a lengthy derivation shown in SM the subsequent heat current pumping into the molecules is
\begin{equation}
\begin{split}
\langle J_1\rangle_{ss} & = \lim_{t\to\infty}\textup{Tr}[\rho_s(t)\tilde{J}_1(t)]\\[0.2cm]
& = (-\gamma)\left[2\bar{\varepsilon}_1\langle a_1^{\dagger}a_1\rangle+\bar{\Delta}\langle a_1^{\dagger}a_2+a_2^{\dagger}a_1\rangle-2\left(E_1n_{\nu_1}^{T_1}\textup{cos}^2\theta+E_2n_{\nu_2}^{T_1}\textup{sin}^2\theta\right)\right]\\[0.2cm]
& = -\frac{\gamma}{6}\bigg\{\left[E_1\left(7\textup{cos}^2\theta-2\textup{sin}\theta\textup{cos}\theta\right)+E_2\left(7\textup{sin}^2\theta+2\textup{sin}\theta\textup{cos}\theta\right)\right]\textup{Y}_1^1\\[0.19cm]
& \qquad\qquad +\left[E_1\left(\textup{cos}^2\theta-2\textup{sin}\textup{cos}\theta\right)+E_2\left(\textup{sin}^2\theta+2\textup{sin}\textup{cos}\theta\right)\right]\left(\textup{Y}_2^2-2\textup{Y}_{12}^{21}\right)\\[0.19cm]
& \qquad\qquad\qquad\qquad\qquad\qquad - 12\left(E_1n_{\nu_1}^{T_1}\textup{cos}^2\theta+E_2n_{\nu_2}^{T_2}\textup{sin}^2\theta\right)\bigg\}
\end{split}
\label{17}
\end{equation}
where $\hbar\nu_1=\frac{1}{2}\left[\bar{\varepsilon}_1+\bar{\varepsilon}_2-\sqrt{(\bar{\varepsilon}_1-\bar{\varepsilon}_2)^2+4\bar{\Delta}^2}\right]$ and $\hbar\nu_2=\frac{1}{2}\left[\bar{\varepsilon}_1+\bar{\varepsilon}_2+\sqrt{(\bar{\varepsilon}_1-\bar{\varepsilon}_2)^2+4\bar{\Delta}^2}\right]$ are the eigenenergies of the coupled oscillators. Moreover the coherence contribution to heat currents $J_1,J_2$ is governed by the term $\langle a_1^{\dagger}a_2+a_2^{\dagger}a_1\rangle$ which will vanishes in secular approximation as shown later. $\langle a_1^{\dagger}a_2\rangle = \sum\limits_{n_1,n_2}\sqrt{n_1n_2}\ \langle n_1-1,n_2|\rho_s|n_1,n_2-1\rangle$. The similar manner for calculating the heat current tranfered by the molecular vibrations and we can simply replace the $\tilde{H}_{int}^{(1)}$ by $\tilde{H}_{int}^{(2)}$ in $J_1$, to reach the following result
\begin{equation}
\begin{split}
\langle J_2\rangle_{ss} = \frac{\gamma}{6}\bigg\{ & \left[E_1\left(\textup{sin}^2\theta-2\textup{sin}\theta\textup{cos}\theta\right)+E_2\left(\textup{cos}^2\theta+2\textup{sin}\theta\textup{cos}\theta\right)\right]\left(\textup{Y}_1^1-2\textup{Y}_{12}^{21}\right)\\[0.19cm]
& + \left[E_1\left(7\textup{sin}^2\theta-2\textup{sin}\theta\textup{cos}\theta\right)+E_2\left(7\textup{cos}^2\theta+2\textup{sin}\theta\textup{cos}\theta\right)\right]\textup{Y}_2^2\\[0.19cm]
& \qquad\qquad\qquad\qquad - 12\left(E_1n_{\nu_1}^{T_2}\textup{sin}^2\theta+E_2n_{\nu_2}^{T_2}\textup{cos}^2\theta\right)\bigg\}
\end{split}
\label{18}
\end{equation}

\begin{figure}
\centering
 $\begin{array}{cc}
  \includegraphics[scale=0.47]{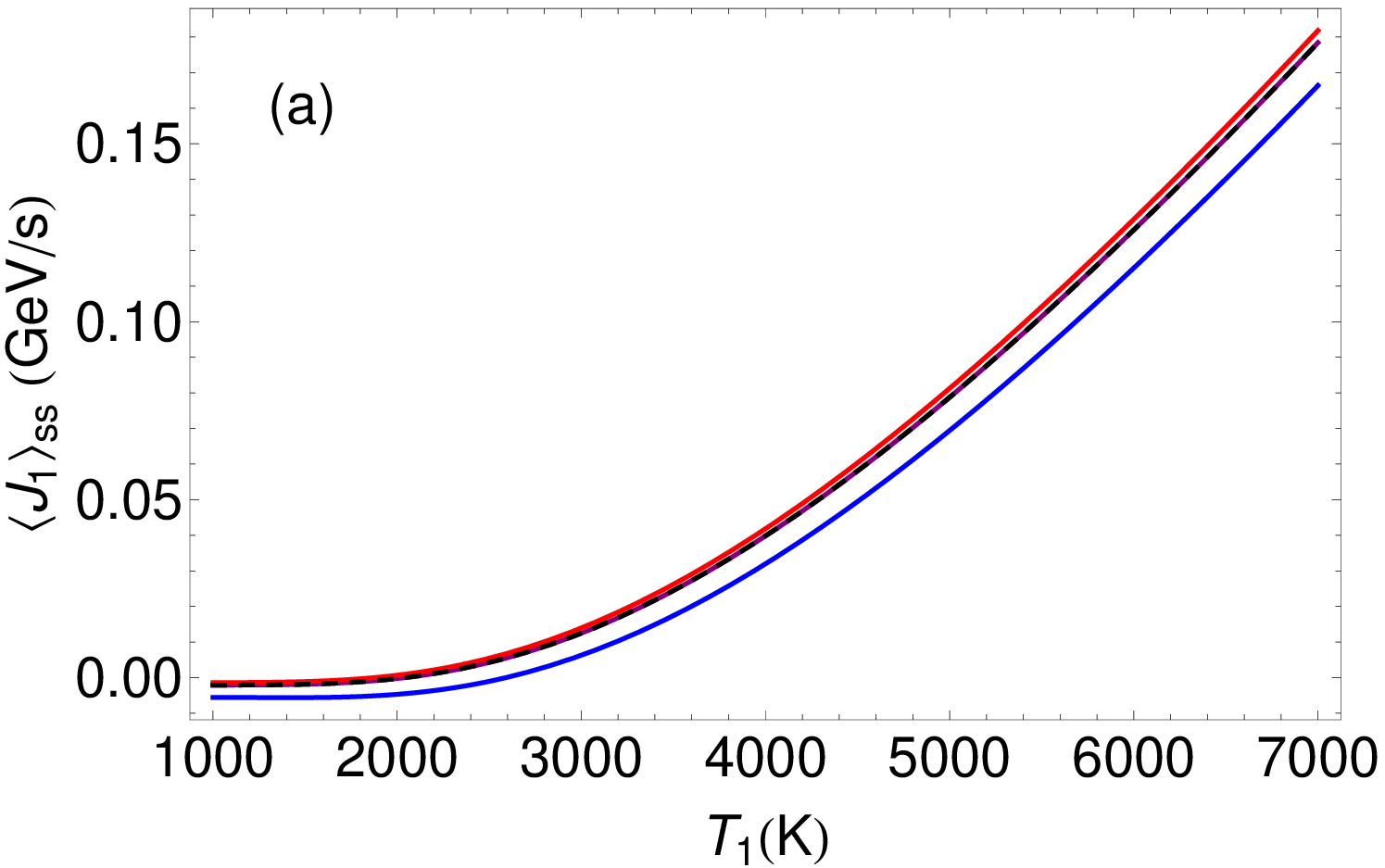}
 &\includegraphics[scale=0.55]{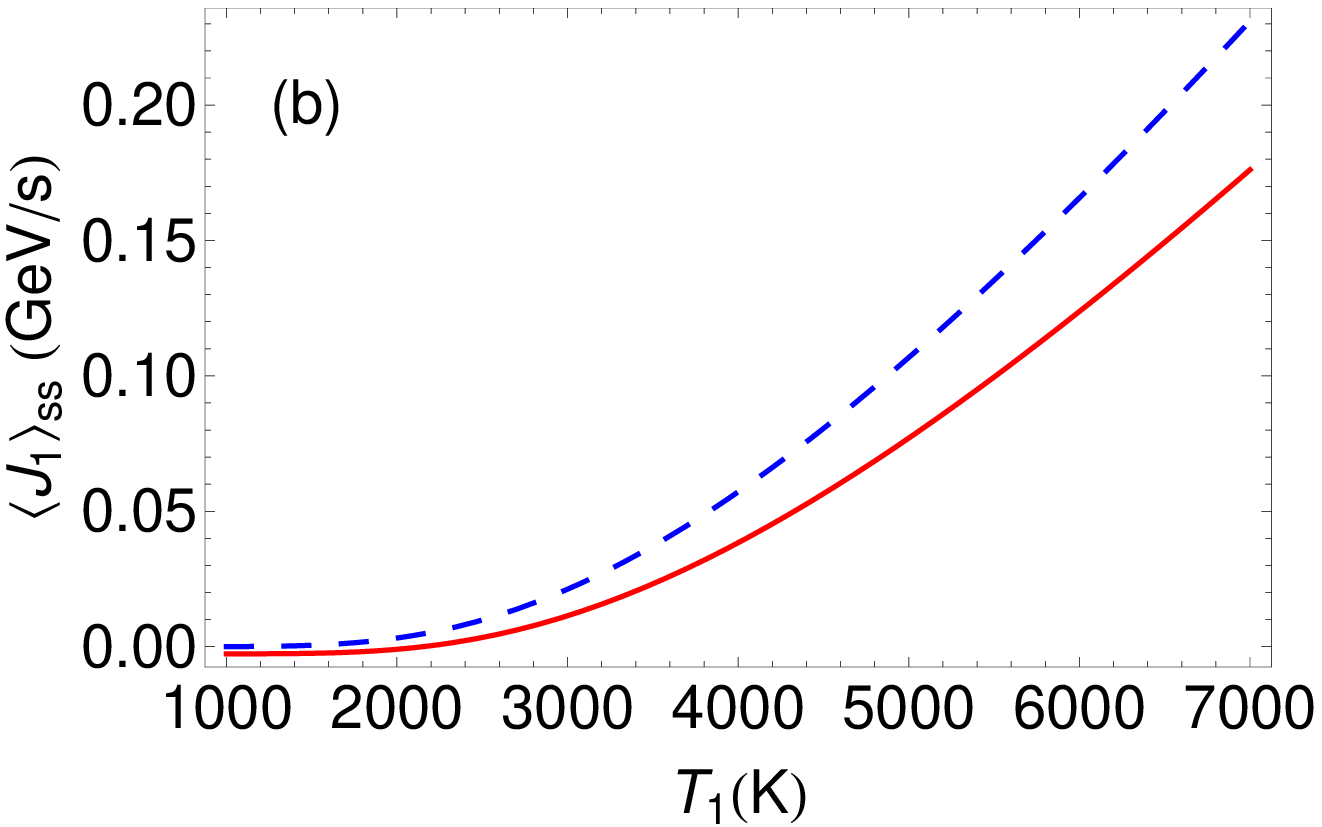}\\
  \includegraphics[scale=0.5]{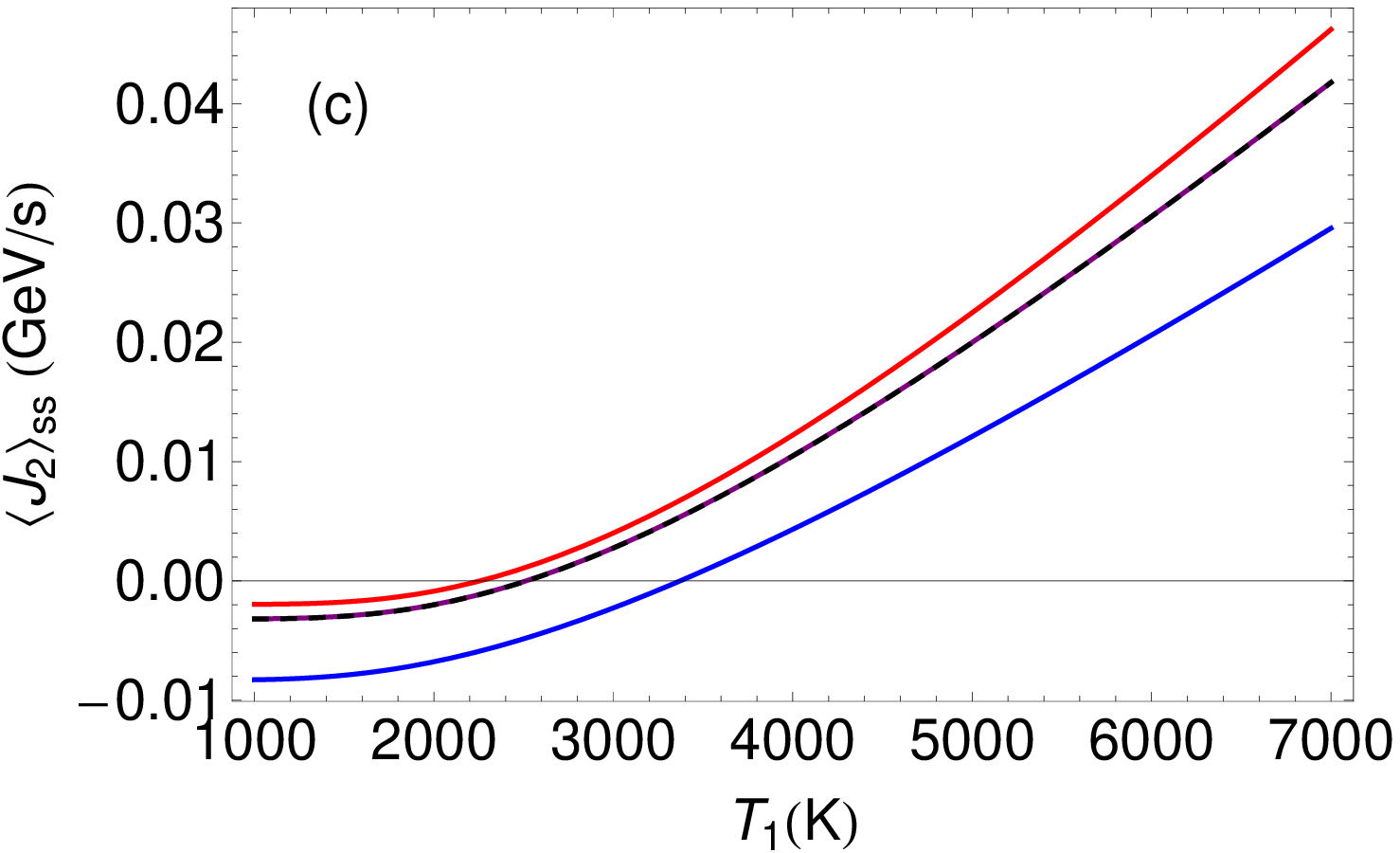}
 &\includegraphics[scale=0.51]{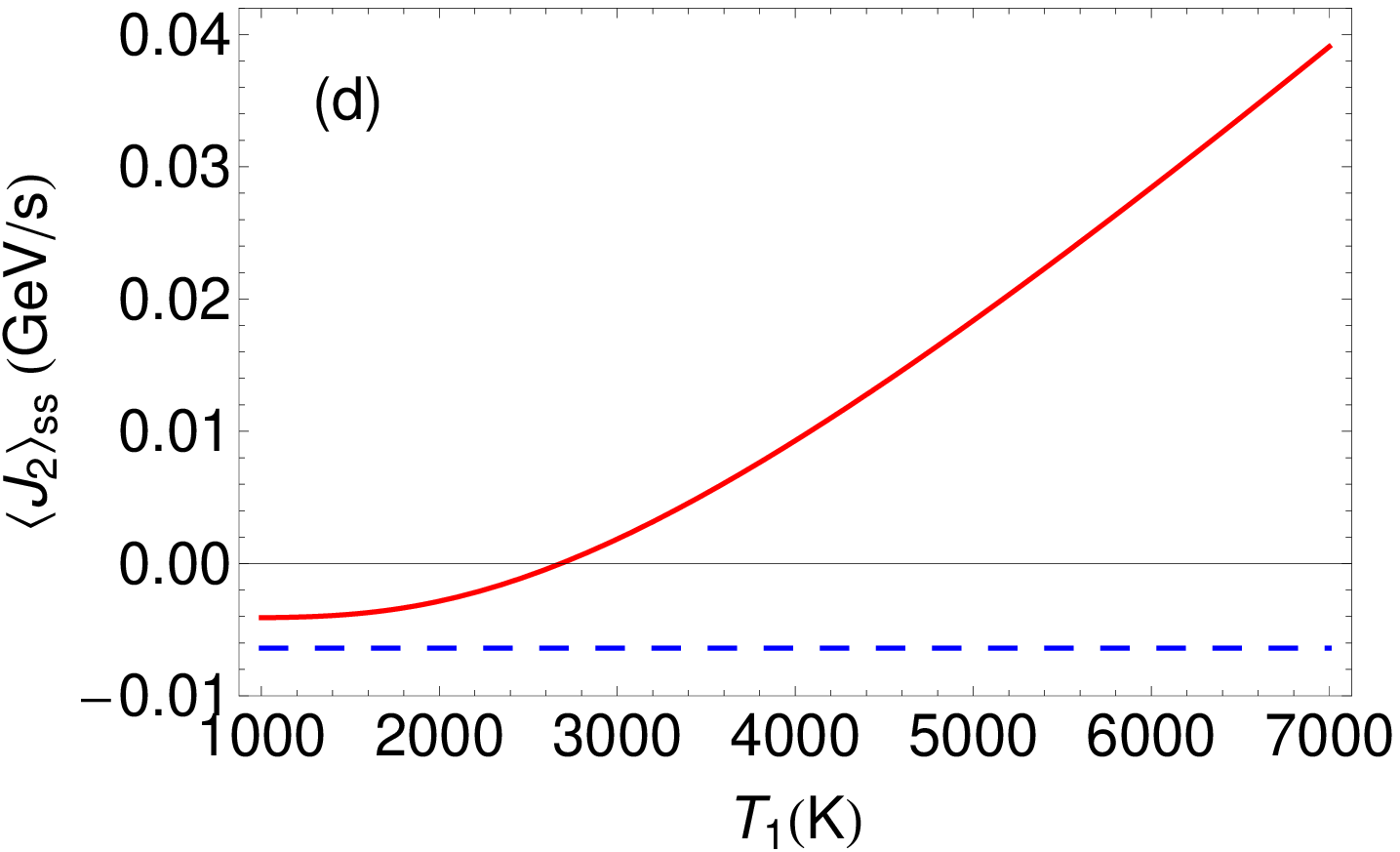}\\
  \includegraphics[scale=0.53]{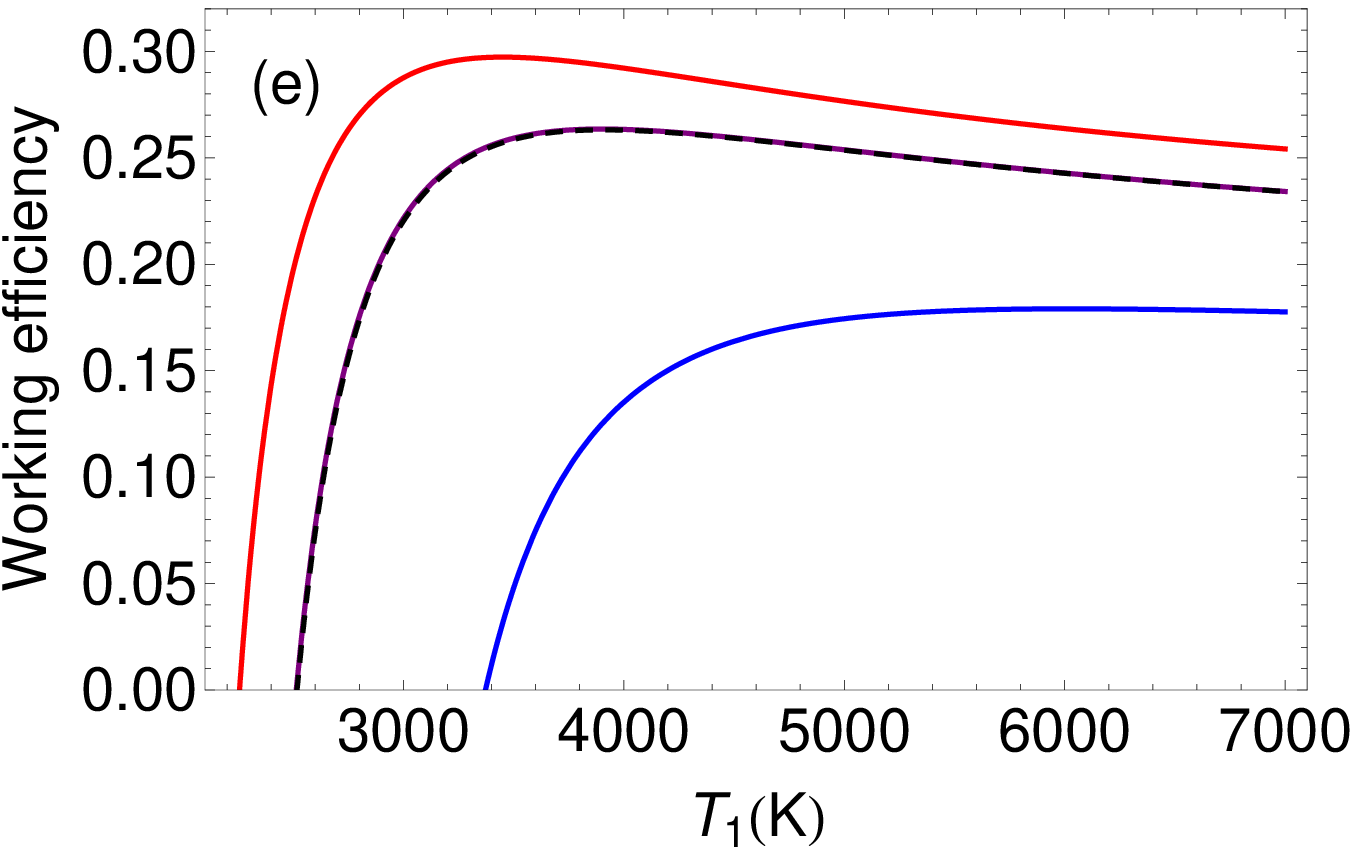}
 &\includegraphics[scale=0.54]{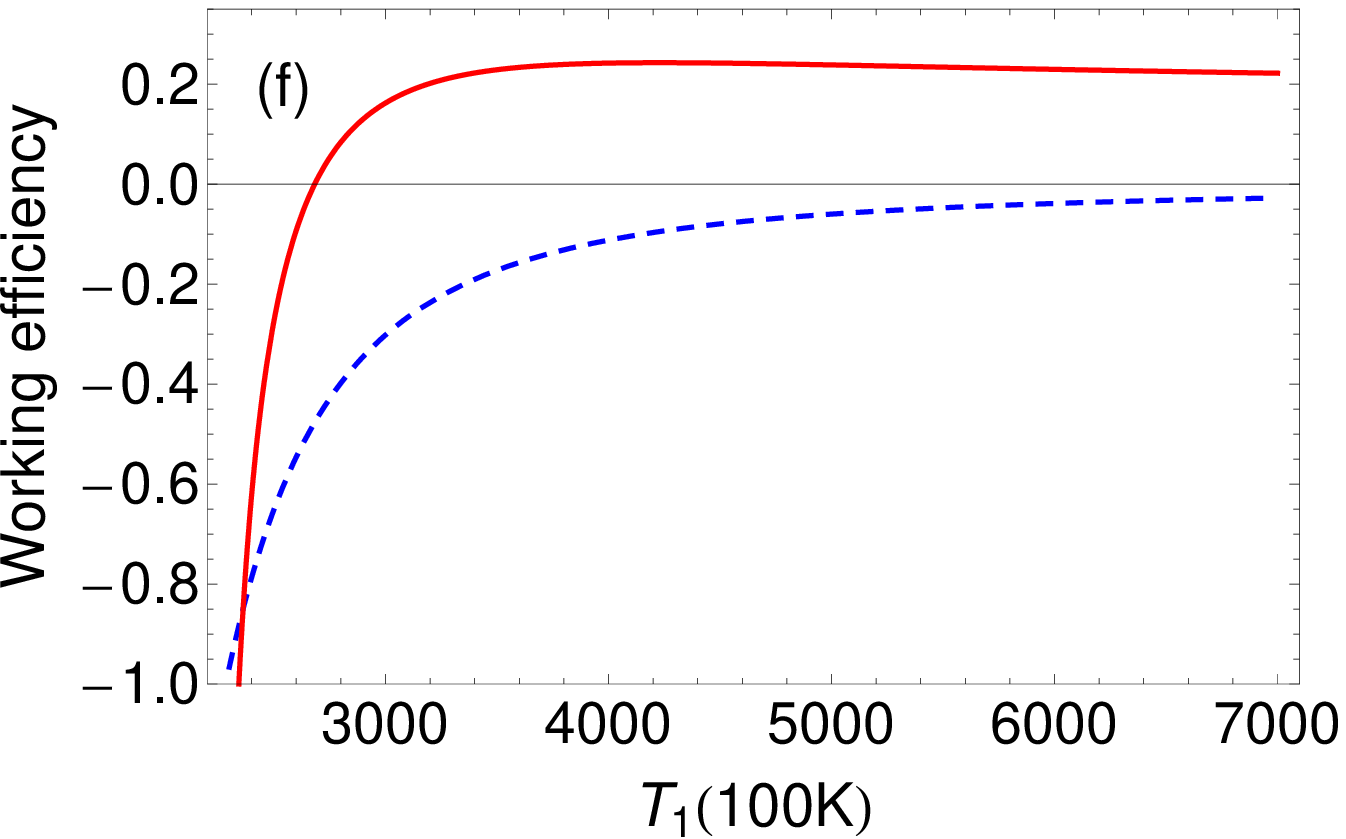}\\
  \end{array}$
\caption{(Color online) (a,b) Energy pumping into molecules from heat source, (c,d) heat current flowing through the molecules into cool reservoir and (e,f) working efficiency vary as a function of $T_1$; Blue, purple and red curves in (a,c,e) correspond to $\delta\bar{\varepsilon}=0.3,\ 0.1,\ 0.01$eV, respectively, where other parameters are $\bar{\Delta}=0.1$eV, $\bar{\varepsilon}_1=1$eV, $T_2=2000$K and $T_3=1000$K. The dashed black curve in (a,c,e) is for $\bar{\Delta}=0.1$eV, $T_3=500$K with other parameters being the same as other curves in (a,c,e); Red and blue (dashed) curves in (b,d,f) correspond to the cases without and with secular approximation, respectively, where the parameters are $\bar{\Delta}=0.1$eV, $\delta\bar{\varepsilon}=0.15$eV, $\bar{\varepsilon}_1=1$eV, $T_2=2000$K and $T_3=1000$K}
\label{heat}
\end{figure}
As a QHE, the stationary working efficiency of the vibratioanl energy transport is naturally defined as $\eta=\frac{\langle J_2\rangle_{ss}}{\langle J_1\rangle_{ss}}$. Fig.\ref{heat}(a) shows the effect of external pumping on the heat flowing into the system according to Eq.(\ref{17}), which demonstrates the improvement of energy pumping into system by the activity of external heat source. 
Since the excitation energy in our system is around 1eV, the considerable excitation with respect to this energy scale needs the effective temperature of the environment to be around 5000 K. On the other hand, the analog is the light-harvesting complex in which the temperature of radiations is around 5800K. Fig.\ref{heat}(c) and \ref{heat}(e) illustrate the heat flow (quantifying the vibrational energy transfer) and working efficiency $\eta$ under the influence of external energy pumping by heat source, according to Eq.(\ref{17}) and (\ref{18}). As seen first, the thermal fluctuation and pumping of the external heat source causes a considerable improvement of the energy transfered by the molecular vibrations and the working efficiency reflected in Fig.\ref{heat}(e) as well. More importantly, it is also shown in Fig.\ref{heat}(c) that the energy transport relates to a critical value of ($\delta\bar{\varepsilon},\ T_1$), under which the vibrational energy transport is suspended. The critical points are determined by acquiring $\langle J_2\rangle_{ss}=0$ which will be shown in Supplementary Information (SI). In Fig.\ref{heat}(c) the critical values of $T_1$ are 3372K, 2514K and 2257K with respect to $\delta\bar{\varepsilon}=0.3,\ 0.1,\ 0.01$eV, respectively. Above the critical point, the promotion of energy current flowing through molecules and the efficiency by thermal fluctuation of the heat source and also the frequency detuning between molecular vibrations can be critically demonstrated by the polarization of the curl flux, illustreated in Fig.\ref{flux}(c,d) and Fig.\ref{flux}(a,b), respectively, since the correlation between the molecular vibrations is enhanced which will be reached later. Therefore we can evidently claim that {\it the curl quantum flux in Eq.(\ref{12}) and (\ref{13}) on microscopic level significantly correlates to and characterises the vibrational energy transport on macroscopic level.} We will come back to this issue when discussing the correlation functions later.

\subsection{Current-current correlation}
\begin{figure}
\centering\includegraphics[scale=0.61]{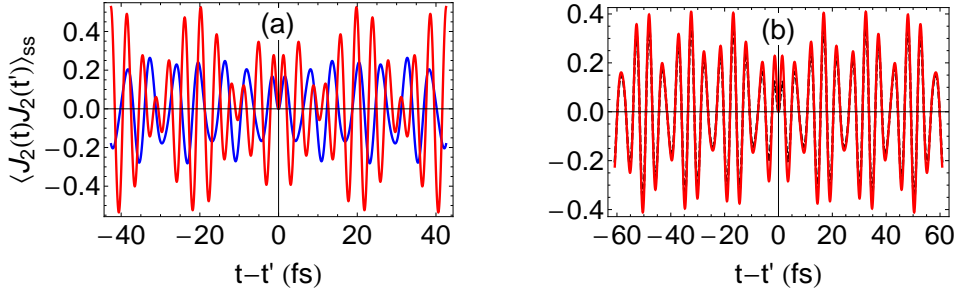}
\caption{(Color online) The smooth part of current-current correlation function as a function of $t-t'$. Red and blue curves correspond to $\delta\bar{\varepsilon}=0.01$ and $0.3$eV, respectively, where other parameters are $\bar{\Delta}=0.1$eV, $T_1=4000$K, $T_2=2000$K and $T_3=1000$K; The black (dashed) curve are for $\delta\bar{\varepsilon}=0.15$eV and $T_1=2000$K, where other parameters are the same.}
\label{JJ_cor}
\end{figure}

To explore the statistical distribution of heat current, one dose not only necessarily calculate the mean heat current as what we did above, but also need to uncover the higher order properties, i.e., the correlations between the currents. 
By replacing the summation over different modes in reservoir by the integration
\begin{equation}
\begin{split}
\sum_{\textbf{k},\sigma} & g_{\textbf{k}\sigma}^2\langle b_{\textbf{k}\sigma}^{(2),\dagger}b_{\textbf{k}\sigma}^{(2)}\rangle e^{i(\nu-\omega_{\textbf{k}\sigma})\Delta t}\\[0.2cm]
& \longrightarrow \gamma\textup{sin}(\nu\Delta t)\int_0^{\infty}\frac{\textup{sin}(\omega_{\textbf{k}\sigma}\Delta t)}{e^{\hbar\omega_{\textbf{k}\sigma}/k T}-1}\textup{d}\omega_{\textbf{k}\sigma}=\frac{\pi\gamma}{2}\textup{sin}(\nu\Delta t)\left(\textup{coth}\frac{\pi\Delta t}{\beta \hbar}-\frac{\beta \hbar}{\pi\Delta t}\right)
\end{split}
\label{rep}
\end{equation}
and after a lengthy calculation we arrive at the current-current correlation function at steady state
\begin{equation}
\begin{split}
& \textup{Re}\left[\langle \tilde{J}_2(t)\tilde{J}_2(t')\rangle_{ss}\right]\\[0.2cm]
& = \frac{\pi\gamma}{\hbar^2}\Bigg\{ \left(E_1^2\langle\eta_1^{\dagger}\eta_1\rangle\textup{sin}^2\theta+E_2^2\langle\eta_2^{\dagger}\eta_2\rangle\textup{cos}^2\theta+E_1E_2\langle\eta_1^{\dagger}\eta_2+\eta_2^{\dagger}\eta_1\rangle\textup{sin}\theta\textup{cos}\theta\right)\delta(\Delta t)\\[0.2cm]
& \quad +\frac{1}{2}\bigg[E_1^2\textup{sin}(\nu_1\Delta t)\langle 2\eta_1^{\dagger}\eta_1+1\rangle\textup{sin}^2\theta+E_2^2\textup{sin}(\nu_2\Delta t)\langle 2\eta_2^{\dagger}\eta_2+1\rangle\textup{cos}^2\theta\\[0.2cm]
& \quad\ +E_1E_2\Big(\textup{sin}(\nu_1\Delta t)+\textup{sin}(\nu_2\Delta t)\Big)\langle\eta_1^{\dagger}\eta_2+\eta_2^{\dagger}\eta_1\rangle\textup{sin}\theta\textup{cos}\theta\bigg]\left(\textup{coth}\frac{\pi\Delta t}{\beta_1\hbar}-\frac{\beta_1\hbar}{\pi\Delta t}\right)\Bigg\}
\end{split}
\label{JJ}
\end{equation}
where $\Delta t\equiv t-t',\ \beta_i\equiv \frac{1}{k T_i}$. Eq.(\ref{JJ}) roughly illustrates the beat oscillation will occur if the detuning between the frequencies of vibrations is suppressed. To demonstrate this, in Fig.\ref{JJ_cor} we show both the cases $\delta\bar{\varepsilon}\ll \bar{\Delta}$ and $\delta\bar{\varepsilon}\gg \bar{\Delta}$ in current-current correlation function (apart from the crusp peak described by the $\delta$-function). The former one characterising the strong vibrational coupling of molecular chain to the molecule stretching (i.e., C=O stretching) reveals the beat oscillation and subsequently a coherent regime of  correlation, in which the correlation is strong. This is in contrast to the latter one characterising the weak vibrational coupling of molecular chain to the molecule stretching, with an incoherent way of correlation in which the correlation is much weaker.

\section{Coherence effect on curl flux and heat transport}
In order to uncover how the entanglement of site-coherence terms to population dynamics gradually affects the nonequilibrium quantites (i.e., curl flux and heat transport), we essentially introduce an adiabatic parameter $\epsilon$ into the coupling coefficients between coherence and populations, which originally as shown in Eq.(\ref{3}) the coherence interacts with the population dynamics. The case $\epsilon=0$ is so-called secular approximation popularly applied to Lindblad equation before \cite{Olaya-Castro08,Lloyd08,Lloyd09,Clark13}, which will be carried out as a comparison to our results with $\epsilon=1$.
\begin{equation}
\begin{split}
\frac{d\rho_s}{dt} = \frac{1}{2\hbar^2} & \bigg\{\sum_{\nu=1}^2\left[\gamma_{\nu}^{T_{\nu},+}\left(a_{\nu}\rho_s a_{\nu}^{\dagger}-a_{\nu}^{\dagger}a_{\nu}\rho_s\right)+\gamma_{\nu}^{T_{\nu},-}\left(a_{\nu}^{\dagger}\rho_s a_{\nu}-a_{\nu}a_{\nu}^{\dagger}\rho_s\right)\right]\\[0.1cm]
& \quad\quad +\sum_{p=1}^2\left[\gamma_p^{T_3,+}\left(a_p\rho_s a_p^{\dagger}-a_p^{\dagger}a_p\rho_s\right)+\gamma_p^{T_3,-}\left(a_p^{\dagger}\rho_s a_p-a_p a_p^{\dagger}\rho_s\right)\right]\\[0.1cm]
& +\epsilon\sum_{\nu\ne p=1}^2\left[\gamma_p^{T_{\nu},+}\left(a_p\rho_s a_{\nu}^{\dagger}-a_{\nu}^{\dagger}a_p\rho_s\right)+\gamma_p^{T_{\nu},-}\left(a_p^{\dagger}\rho_s a_{\nu}-a_{\nu}a_p^{\dagger}\rho_s\right)\right]\\[0.1cm]
&  +\epsilon\sum_{j\ne p=1}^2\left[\gamma_p^{T_3,+}\left(a_p\rho_s a_j^{\dagger}-a_j^{\dagger}a_p\rho_s\right)+\gamma_p^{T_3,+}\left(a_p^{\dagger}\rho_s a_j-a_j a_p^{\dagger}\rho_s\right)\right]
\bigg\}+\textup{h.c.}
\end{split}
\label{24}
\end{equation}
in which the $\epsilon$-terms describe the behaviors of the dynamics when the coherence adiabatically comes into the system. Correspondingly the dynamical equation in the coherent representation is of the form
\begin{equation}
\begin{split}
\frac{\partial}{\partial t} & P(\alpha_{\beta},\alpha_{\beta}^*) = \gamma \left[2\left(\frac{\partial}{\partial\alpha_1}\alpha_1+\frac{\partial}{\partial\alpha_2}\alpha_2\right)+\epsilon\left(\frac{\partial}{\partial\alpha_1}\alpha_2+\frac{\partial}{\partial\alpha_2}\alpha_1\right)+\textup{c.c.}\right]P(\alpha_{\beta},\alpha_{\beta}^*)\\[0.26cm]
& \ \quad +\gamma\left[2\textup{Y}_1^1\frac{\partial^2}{\partial\alpha_1^*\partial\alpha_1}+2\textup{Y}_2^2\frac{\partial^2}{\partial\alpha_2^*\partial\alpha_2}+\epsilon\textup{Y}_{12}^{21}\left(\frac{\partial^2}{\partial\alpha_1^*\partial\alpha_2}+\frac{\partial^2}{\partial\alpha_1\partial\alpha_2^*}\right)\right]P(\alpha_{\beta},\alpha_{\beta}^*)
\end{split}
\label{25}
\end{equation}
thus the drift and diffusion matrices are (in the order of $\{1,\ 2,\ 1^*,\ 2^*\}$)
\begin{equation}
\begin{split}
\Sigma = \gamma\begin{pmatrix}
                 2+\eta & \epsilon & 0 & 0\\[0.12cm]
                 \epsilon & 2-\eta & 0 & 0\\[0.12cm]
                 0 & 0 & 2+\eta & \epsilon\\[0.12cm]
                 0 & 0 & \epsilon & 2-\eta\\
                \end{pmatrix}
,\quad D = \gamma\begin{pmatrix}
                  0 & 0 & \textup{Y}_1^1 & \frac{\epsilon}{2}\textup{Y}_{12}^{21}\\[0.12cm]
                  0 & 0 & \frac{\epsilon}{2}\textup{Y}_{12}^{21} & \textup{Y}_2^2\\[0.12cm]
                  \textup{Y}_1^1 & \frac{\epsilon}{2}\textup{Y}_{12}^{21} & 0 & 0\\[0.12cm]
                  \frac{\epsilon}{2}\textup{Y}_{12}^{21} & \textup{Y}_2^2 & 0 & 0\\
                 \end{pmatrix}
\end{split}
\label{26}
\end{equation}
where $\eta$ is a small and positive number, which ensures the uniqueness of the solution to Eq.(\ref{25}). Finally we will carry out the limit $\eta\rightarrow 0^+$. By introducing $f_{\pm}=\sqrt{1+\frac{\eta^2}{\epsilon^2}}\pm \frac{\eta}{\epsilon}$, the steady-state solution to Eq.(\ref{25}) can be written as $P_{ss}=\frac{\tilde{a}\tilde{b}-\tilde{c}^2}{\pi^2}\textup{exp}\{-[\tilde{a}|\alpha_1|^2+\tilde{b}|\alpha_2|^2+2\tilde{c}\textup{Re}(\alpha_1^*\alpha_2)]\}$ where
\begin{equation}
\begin{split}
& \tilde{a} = \frac{\textup{A}_{11}^{24}\textup{Y}_1^1+\textup{A}_{22}^{24}\textup{Y}_2^2+\textup{A}_{1221}^{24}\textup{Y}_{12}^{21}}{\textup{det}(B)},\ \ 
\tilde{b} = \frac{\textup{A}_{11}^{13}\textup{Y}_1^1+\textup{A}_{22}^{13}\textup{Y}_2^2+\textup{A}_{1221}^{13}\textup{Y}_{12}^{21}}{\textup{det}(B)},\\[0.2cm]
& \tilde{c} = -\frac{\textup{A}_{11}^{14}\textup{Y}_1^1+\textup{A}_{22}^{14}\textup{Y}_2^2+\textup{A}_{1221}^{14}\textup{Y}_{12}^{21}}{\textup{det}(B)}\\[0.2cm]
& B=\begin{pmatrix}
     \textup{A}_{11}^{13}\textup{Y}_1^1+\textup{A}_{22}^{13}\textup{Y}_2^2+\textup{A}_{1221}^{13}\textup{Y}_{12}^{21} & \textup{A}_{11}^{14}\textup{Y}_1^1+\textup{A}_{22}^{14}\textup{Y}_2^2+\textup{A}_{1221}^{14}\textup{Y}_{12}^{21}\\[0.2cm]
     \textup{A}_{11}^{14}\textup{Y}_1^1+\textup{A}_{22}^{14}\textup{Y}_2^2+\textup{A}_{1221}^{14}\textup{Y}_{12}^{21} & \textup{A}_{11}^{24}\textup{Y}_1^1+\textup{A}_{22}^{24}\textup{Y}_2^2+\textup{A}_{1221}^{24}\textup{Y}_{12}^{21}\\
    \end{pmatrix}
\end{split}
\label{27}
\end{equation}
where the expressions of $A_{...}$ are provided in Appendix C. Since $\lim\limits_{\eta\to 0}\lim\limits_{\epsilon\to 1}f_{\pm}=1$ and $\lim\limits_{\eta\to 0}\lim\limits_{\epsilon\to 0}f_+=+\infty,\ \lim\limits_{\eta\to 0}\lim\limits_{\epsilon\to 0}f_-=0$, it is straightforward to verify that Eq.(\ref{27}) will reduce to Eq.(\ref{10}) and $\bar{a},\ \bar{b}$ in the forthingcoming $P_{sec}^{ss}$, respectively. Therefore the heat currents tranfered by molecular vibrations is

\begin{figure}
\centering
 $\begin{array}{cc}
  \includegraphics[scale=0.485]{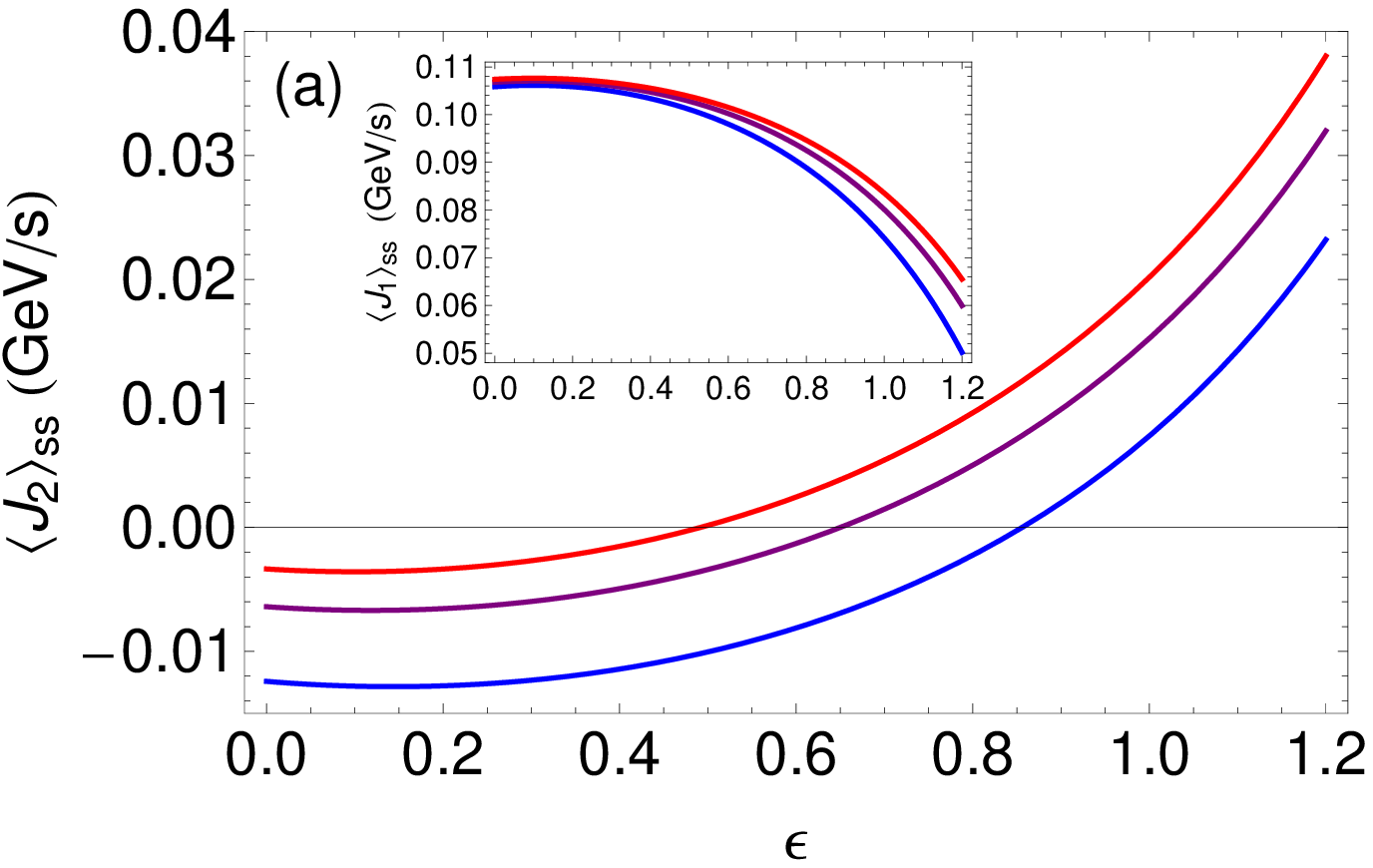}
 &\includegraphics[scale=0.51]{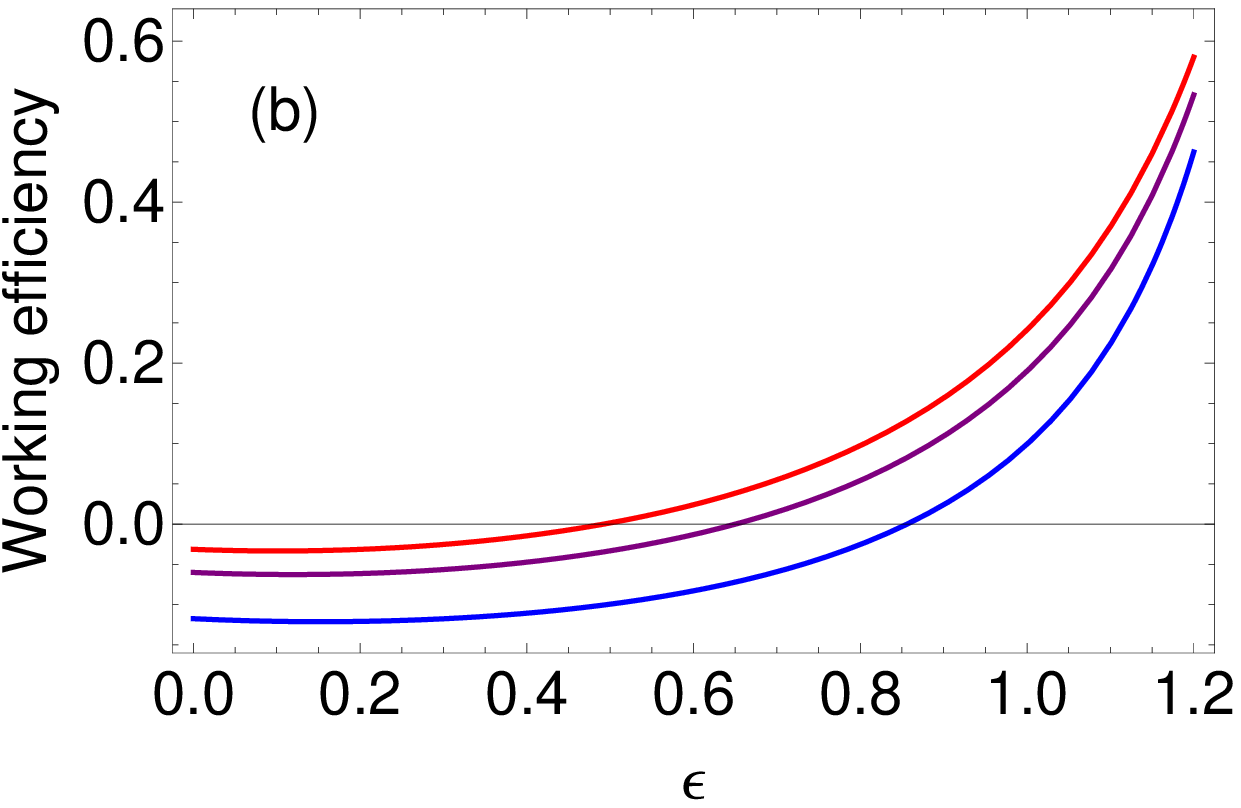}\\
  \end{array}$
\caption{(Color online) The contribution of coherence on heat transport. (a) The heat flow pumping from the heat source (small) as well as the heat current intermediated by molecular vibrations (large), (b) working efficiency of the QHE. The blue, purple and red lines are for $\delta\bar{\varepsilon}=0.3$, $0.15$ and $0.01$eV, respectively. Other parameters are $\bar{\Delta}=0.1$eV, $T_1=5000$K, $T_2=2000$K, $T_3=1000$K and $\eta=0.0001$.}
\label{coh}
\end{figure}

\begin{equation}
\begin{split}
\langle J_2\rangle_{ss} = 2 & \gamma\bigg\{ \left[E_1\left(\textup{A}_{11}^{24}\textup{sin}^2\theta+\textup{A}_{11}^{14}\textup{sin}\theta\textup{cos}\theta\right)+E_2\left(\textup{A}_{11}^{24}\textup{cos}^2\theta-\textup{A}_{11}^{14}\textup{sin}\theta\textup{cos}\theta\right)\right]\textup{Y}_1^1\\[0.2cm]
& \quad +\left[E_1\left(\textup{A}_{22}^{24}\textup{sin}^2\theta+\textup{A}_{22}^{14}\textup{sin}\theta\textup{cos}\theta\right)+E_2\left(\textup{A}_{22}^{24}\textup{cos}^2\theta-\textup{A}_{22}^{14}\textup{sin}\theta\textup{cos}\theta\right)\right]\textup{Y}_2^2\\[0.2cm]
& +\left[E_1\left(\textup{A}_{1221}^{24}\textup{sin}^2\theta+\textup{A}_{1221}^{14}\textup{sin}\theta\textup{cos}\theta\right)+E_2\left(\textup{A}_{1221}^{24}\textup{cos}^2\theta-\textup{A}_{1221}^{14}\textup{sin}\theta\textup{cos}\theta\right)\right]\textup{Y}_{12}^{21}\\[0.2cm]
& \qquad\qquad\qquad\qquad\qquad\qquad - E_1n_{\nu_1}^{T_2}\textup{sin}^2\theta - E_2n_{\nu_2}^{T_2}\textup{cos}^2\theta\bigg\}
\end{split}
\label{30}
\end{equation}
and the heat current pumping by the thermal fluctuations in heat source can be reached by the replacement $\textup{cos}\theta\rightarrow\textup{sin}\theta,\ \textup{sin}\theta\rightarrow -\textup{cos}\theta$ in the expression of $J_2$. The density-density correlation function is
\begin{equation}
\begin{split}
\tilde{C}_4 = 
\frac{\tilde{e}^4\textup{sin}^2\beta}{4(1-\tilde{e}^2)+\tilde{e}^4\textup{sin}^2\beta},\ \ \tilde{e}=\sqrt{\frac{2\sqrt{(\tilde{a}-\tilde{b})^2+4\tilde{c}^2}}{\tilde{a}+\tilde{b}+\sqrt{(\tilde{a}-\tilde{b})^2+4\tilde{c}^2}}}
\end{split}
\label{31}
\end{equation}

Now within the secular approximation, the entanglement between coherence and populations in Eq.(\ref{25}) is mathematically reflected by the mixed differentials with respect to $\alpha_1,\alpha_2$ which disappear. Then the steady-state solution is simplified to $P_{sec}^{ss}(\alpha_{\beta},\alpha_{\beta}^*) = \frac{\bar{a}\bar{b}}{\pi^2}\  e^{-(\bar{a}|\alpha_1|^2+\bar{b}|\alpha_2|^2)},\quad\textup{with}\quad
\bar{a} = \frac{2}{\textup{Y}_1^1},\ \bar{b} = \frac{2}{\textup{Y}_2^2}$. By reducing to the ($x_1,x_2$) space, the polarization of curl flux is along the $x_1$-axis, since
\begin{equation}
\begin{split}
\lim\limits_{\eta\rightarrow 0}\lim\limits_{\epsilon\rightarrow 0}\beta =0\ \Rightarrow\ \lim\limits_{\eta\rightarrow 0}\lim\limits_{\epsilon\rightarrow 0}\tilde{C}_4 = 0
\end{split}
\label{C4}
\end{equation}
as firstly mentioned before. Therefore the secular approximation leads to the death of correlation between molecular vibrations (numerical verification refers to Fig.2 in Appendix C).

Eq.(\ref{31}), Eq.(\ref{C4}), Fig.\ref{correlation}(c,d) and Fig.\ref{correlation}(f) illustrate the coherence effect on the curl quantum flux that the site-basis coherence causes the correlations between the molecular vibrations, which is determined by the polarization and orientation of the curl flux in coherence space, as uncovered before. Hence the shape and orientation of flux provides a quantification to the coherence contribution to nonequilibrium-related quantities. 

As is shown in Fig.\ref{heat}(b,d,e), the coherence terms considerably contribute to heat current flowing through the molecules and the efficiency for the vibrational energy transport. This is because of the promotion of density-density correlation originated from the coherence quantified by the polarization and orientation of the curl flux, as illustrated in Eq.(\ref{31}), (\ref{C4}) and Fig.\ref{correlation}(c,d) and Fig.\ref{correlation}(f) before. Consequently it is evident that the site-basis coherence is critical and non-trivial for the nonequilibrium behaviors on microscopic level and the quantum transport on macroscopic level. Experimentally this coherence effect can be observed by the high-resolution multidimensional laser spectroscopy, as being applied in the study of long-lived coherence in photosynthesis. On the other hand, the vibrational interfacial energy transfer can be investigated using surface-specific 2D-IR sum-frequency generation (2D-SFG) spectroscopy, by studying the effect of widely tunable excitation pulses (2100-3000cm$^{-1}$) in heavy water (D$_2$O) \cite{Bonn11}. 

As is shown in Fig.\ref{coh}(a), the heat pumping into the molecules by the thermal fluctuations in the heat source (small figure) is suppressed while the heat transfered by the molecular vibrations (large figure) is considerably improved, as the coupling between site-coherence and population dynamics increases adiabatically. This results in the significant promotion of working efficiency, as illustated in Fig.\ref{coh}(b). More importantly, it is noted that {\it the slight increase of couling between coherence and population dynamics beyond $\epsilon=1$ causes a surprising improvement of efficiency to ${\it \ge 42\%}$, in comparison to the result in Fig.\ref{heat}(f) (blue line). This demonstrates the possibility for optimizing the quality of QHE in the regime of large interaction with coherence.} Fig.\ref{correlation}(d) and \ref{correlation}(f) support the conclusion in above that the correlation between the occupations in different molecules is generated and further strengthened by the contribution of coherence, as quantified by Eq.(\ref{31}).

\section{Conclusion and remarks}
In this work we develope a theoretical framework of curl quantum flux in continous space, to study the microscopic nonequilibrium behaviors and the macroscopic vibrational energy transport in molecules. It was found in an analytical manner of the connection between the microscopic curl flux and the macroscopic quantum transport (i.e., correlation function) from the geometric and magnitude perspectives. By adiabatically tuning the coupling of site-coherence to the population dynamics and further comparing to the secular approximation, the coherence is demonstrated to be essential to generate the density-density correlations and further facilitate the heat transport process by a considerable improvement. These {\it non-trivial} coherence effects originate from the microscopic channel, in which coherence-population entanglement results in the slender-cigar shape of curl flux polarized in the vicinity of anti-diagonal, quantified by its geometric parameters. By exploring the current-current correlation, 
 the beat oscillation feature with strong correlation charaterises the delocalization of vibrations induced by large VP coupling, contrary to weak VP coupling where the vibrations are kept localized. Our investigation provides the possibility of probing the microscopic quantum flux in experiments and also the insights for the exploration of the nonequilibrium heat transport in more general quantum systems, i.e., the molecular chain with several vibrational modes.

\section{Acknowledgements}
We thank the support from the grant NSF-MCB-0947767.


\end{document}